\documentclass[journal]{IEEEtran}
\usepackage{amsmath}
\usepackage{amssymb}
\usepackage{amsthm}
\usepackage{cite}
\usepackage{color}
\usepackage{xcolor}
\usepackage{caption}
\usepackage{epsfig,latexsym}
\usepackage{float}
\usepackage{fancyhdr}
\usepackage{graphicx}
\usepackage{indentfirst}
\usepackage{mdwmath}
\usepackage{mdwtab}
\usepackage{subfigure}
\usepackage{setspace}
\usepackage{times}
\usepackage{url}
\usepackage{multirow}
\usepackage{algorithm}
\usepackage{algorithmic}
\usepackage{verbatim}
\usepackage{algorithm,algorithmic}
\usepackage{mathrsfs}

\newtheorem{remark}{Remark}
\newtheorem{theorem}{Theorem}

\newtheorem{lemma}{Lemma}

\newtheorem{corollary}{Corollary}

\begin{document}

\title{{Integrated Positioning and Communications for PASS: A Robust Approach}}

\author{Yaoyu Zhang, Xin Sun, Jun Wang, Tianwei Hou,~\IEEEmembership{Member,~IEEE}, Anna Li,~\IEEEmembership{Member,~IEEE},\\ 
Yuanwei Liu,~\IEEEmembership{Fellow,~IEEE}, and Arumugam Nallanathan,~\IEEEmembership{Fellow,~IEEE}

\thanks{This work was supported in part by the Fundamental Research Funds for the Central Universities under Grant 2023JBZY012 and 2024JBMC014, in part by the National Natural Science Foundation for Young Scientists of China under Grant 62201028, in part by Young Elite Scientists Sponsorship Program by CAST under Grant 2022QNRC001, in part by the Beijing Natural Science Foundation L232041, and in part by EPSRC grant numbers to acknowledge are EP/W004100/1, EP/W034786/1 and EP/Y037243/1.}       
\thanks{Yaoyu Zhang,  Xin Sun, Jun Wang and Tianwei Hou are with the School of Electronic and Information Engineering, Beijing Jiaotong University, Beijing 100044, China (e-mail: 24110070@bjtu.edu.cn; xsun@bjtu.edu.cn; wangjun1@bjtu.edu.cn; twhou@bjtu.edu.cn). (Corresponding author: Tianwei Hou.)}  
\thanks{Anna Li is with the School of Computing and Communications, Lancaster University, Lancaster LA1 4WA, U.K. (e-mail: a.li16@lancaster.ac.uk). }
\thanks{Yuanwei Liu is with the Department of Electrical and Electronic Engineering, The University of Hong Kong, Hong Kong (e-mail: yuanwei@hku.hk). } 
\thanks{Arumugam Nallanathan is with the School of Electronic Engineering and Computer Science, Queen Mary University of London, London E1 4NS, U.K., and also with the Department of Electronic Engineering, Kyung Hee University, Yongin-si, Gyeonggi-do 17104, Korea (e-mail: a.nallanathan@qmul.ac.uk).}}

\maketitle

\begin{abstract}
The pinching-antenna systems (PASS), which dynamically activate and relocate the pinching-antennas (PAs) along the dielectric waveguide, offer unprecedented potential for integrated positioning and communication. The multi-waveguide-based uplink positioning approaches for indoor environments are first proposed in this paper, and the downlink communication performance is analyzed. Two possible scenarios, multi-waveguide single-PA (MWSP) and multi-waveguide multi-PA (MWMP), are considered under the assumptions of line-of-sight channels and a single, stationary user. For the MWSP scenario, the received signal strength indication (RSSI)-based ranging method and the MWSP-based least square (LS) positioning algorithm are developed. To gain deeper insights, a comprehensive error analysis of the LS positioning algorithm is conducted. Subsequently, for the MWMP scenario, the closed-form expression of the superposed signal is derived. According to the signal power, the MWMP-based grid search algorithm is proposed and the estimation error of proposed algorithm is analyzed. Then, based on the user's positioning result, the PAs are relocated to provide downlink communication service, and the achievable data rate of MWSP and MWMP scenarios are analyzed. Numerical results validate the correctness of our analysis, which show that: i) For the MWSP scenario, a smaller geometric dilution of precision (GDoP) leads to a lower average positioning error. Furthermore, even when the GDoP is large, the regions where the distances to PAs are nearly equal achieve the best accuracy. ii) For the MWMP scenario, non-parallel waveguide deployment improves positioning accuracy, although errors increase with the number of PAs. iii) The noise has a serious double-impact on data rate. There is a trade-off between positioning accuracy and communication performance.
\end{abstract}


\section{Introduction}

The sixth-generation (6G) wireless communication network is expected to become a unifying infrastructure that enables ubiquitous, seamless, and intelligent interconnection among humans, machines, and objects, thereby driving unprecedented requirements for network performance. To meet these requirements, massive multiple-input multiple-output (MIMO) technology has emerged as a cornerstone for improving spectral efficiency and link reliability by leveraging the spatial dimension~\cite{MIMO_1},~\cite{MIMO_2}. However, traditional MIMO systems rely on fixed antenna arrays, whose static deployment has certain limitations in the complex and dynamic wireless environments. The fixed antennas are easily blocked by obstacles, which causes communication links to degrade from high-quality line-of-sight (LoS) transmission to significantly lower performance non-line-of-sight (NLoS) transmission, making it difficult to adapt to user mobility and dynamic environmental changes~\cite{MIMO_3},~\cite{MIMO_4}.

To overcome the performance limitations imposed by fixed antennas, researchers have investigated reconfigurable antenna technologies. Among these technologies, reconfigurable intelligent surface (RIS) technology offers a promising solution by introducing a new dimension for optimizing signal propagation, which is achieved through intelligent control of electromagnetic wave reflections~\cite{RIS_1},~\cite{RIS_2}. Building on this concept, flexible-antenna systems have recently emerged, which aim to enhance dynamic adaptability by enabling physical movement or reconfiguration of antennas~\cite{Flexible_antennas,flexible_2,flexible_3}. Although techniques such as movable and fluid antennas can mitigate small-scale fading by adjusting antenna positions at wavelength-level scales~\cite{movable_antennas},~\cite{fluid_antennas}, they are still unable to effectively counter the large-scale fading due to physical obstructions and cannot restore LoS connectivity once the link has been disrupted. 

As a promising technique for overcoming blockage and restoring LoS connections in dynamic wireless environments, pinching-antenna (PA) systems (PASS) have been proposed~\cite{PASS_1,PASS_2,PASS_3,PASS_4}. The PASS architecture mainly consists of a dielectric waveguide and multiple dielectric elements pinched by the plastic clamps, i.e., PAs. In practical deployments, the low-loss dielectric waveguide is connected to the access point (AP) and installed along the ceiling boundary. By pinching the PA at an arbitrary location along the waveguide, a radiating element can be formed at that point~\cite{PA}. From a practical implementation perspective, this pinching mechanism can be realized either mechanically or automatically. In the initial prototype, the PAs are manually attached using plastic pinches to provide temporary coverage~\cite{PA}. For dynamic indoor environments, the PAs can be mounted on motorized sliders running parallel to the waveguide, thereby enabling real-time repositioning controlled by the AP~\cite{PASS_3}. Driven by this automated mechanism, PAs can dynamically move along the waveguide to positions near users, thereby effectively bypassing obstacles and transforming wireless communication from the traditional ``last mile'' into an almost wired-like ``last meter'' connection~\cite{PASS_3,PASS_4}. Furthermore, PASS is economically viable and scalable. Unlike conventional active antenna arrays, which require dedicated and expensive radio-frequency (RF) chains and phase shifters for each element, PASS utilizes low-cost dielectric materials and entirely passive PAs. Multiple PAs can share a single RF chain connected to the waveguide, which significantly reduces hardware complexity, deployment cost, and power consumption, making PASS a highly scalable solution for realistic indoor deployments~\cite{PASS_4}.

Recently, many researchers have explored the performance analysis and optimization of PASS. Wang et al. investigated a non-orthogonal multiple access (NOMA) assisted downlink PA system, which can activate multiple PAs at predefined locations on a dielectric waveguide to support users~\cite{PA_apply1}. Ouyang et al. presented a closed-form characterization of the array gain limit in PASS and explored its variation with respect to antenna spacing. Simulation results show that appropriate configuration of antenna quantity and inter-element spacing allows PASS to deliver superior array gain compared with traditional antenna deployments~\cite{PA_perfor1}. Ding et al. investigated the impact of LoS blockage on PASS, indicating that such blockage leads to greater performance benefits for PASS compared to traditional antenna systems.~\cite{LoS_blockage1}. Wang et al. further explored the capability of establishing LoS links under blockage by formulating a matching-based optimization framework, demonstrating that the proposed approach can dynamically restore LoS connectivity and significantly improve system throughput~\cite{LoS_blockage2}. Furthermore, Xu et al. analyzed the effect of the waveguide attenuation in LoS blockage scenario, and a dynamic sample average approximation algorithm was proposed to maximize the average rate of users~\cite{LoS_blockage3}. In addition, Zhou et al. proposed a gradient-based meta-learning joint optimization algorithm, in which the joint optimization of beamforming design and antenna deployment was performed to maximize the weighted sum rate~\cite{optimize1}. Sun et al. proposed an element-wise sequential optimization approach with low computational burden for determining PA locations in the waveguide, leading to enhanced throughput across both uplink and downlink channels of PASS~\cite{optimize2}. Gan et al. developed a NOMA-assisted PASS framework, in which the transmit power reduction is realized by the integrated design of transmit beamforming, pinching beamforming, and power control~\cite{optimize3}. 

Despite the promising performance of PASS in enhancing communication efficiency, its potential for indoor positioning remains largely unexplored. Existing indoor positioning methods generally utilize wireless fidelity (Wi-Fi), ultra wide band (UWB) and fixed antenna arrays. He et al. provided a comprehensive survey on Wi-Fi fingerprinting techniques and system deployment~\cite{wifi}. However, as noted in the literature, Wi-Fi-based positioning methods cannot adapt to dynamic channel environments and suffer from high computational complexity. Alarifi et al. presented an overview of advanced positioning approaches, including a detailed comparative analysis of UWB positioning technologies~\cite{UWB}. Despite the high accuracy of UWB-based positioning, it requires specialized, costly hardware infrastructure. Wielandt et al. investigated the performance of angle-of-arrival-based indoor positioning systems, with a particular focus on the positioning errors of 2.4-GHz linear antenna arrays under various environments~\cite{AOA}. Recently, many researchers have proposed the RIS-assisted indoor positioning systems. Pu et al. built a virtual LoS link through RIS, which improved the positioning performance in NLoS environments~\cite{RIS_indoor1}. Ma et al. applied RIS to indoor positioning with UWB technique and derived the Cramér–Rao limit for the proposed localization approach~\cite{RIS_indoor2}. Xu et al. designed a machine learning-enhanced indoor positioning method based on RIS-assisted channel configuration and analysis, which is suitable for large-scale warehouse applications~\cite{RIS_indoor3}. Importantly, zhang et al. proposed a PASS-based indoor positioning approach and developed a positioning algorithm based on a single waveguide with multi-PA, where the proposed method is efficient and particularly applicable to PASS~\cite{PA_indoor}.

\subsection{Motivation and contribution}

Unlike the traditional fixed antenna systems that are vulnerable to blockage, PASS actively reconfigure antenna positions to maintain LoS propagation. By sustaining the short-range ``last-meter'' links, PASS can achieve superior signal quality and energy efficiency, which offers distinct advantages in complex indoor environments. However, based on the previous analysis, it is evident that the majority of studies have concentrated on PASS for communications, whereas the research on indoor positioning remains relatively unexplored. Existing studies on PASS generally assume that user locations are known when optimizing PA positions, leaving the fundamental challenge of acquiring positioning information unaddressed. In particular, single-waveguide-based indoor positioning methods suffer from the limitation that only one PA can be activated in each time slot, resulting in poor real-time positioning performance. Moreover, indoor positioning approaches for multi-waveguide PASS have not yet been explored. Even in current uplink multi-waveguide schemes, each waveguide is typically restricted to a single PA, and the critical issue of uplink signal superposition under simultaneous multi-PA activation remains unresolved.

To overcome these limitations, this paper investigates the comprehensive uplink positioning techniques for multi-waveguide PASS and develops dedicated positioning algorithms for both multi-waveguide single-PA (MWSP) and multi-waveguide multi-PA (MWMP) scenarios. Through these approaches, we establish the PASS-based integrated positioning and communication architecture. It is important to note that unlike waveform-level integration, the ``integration'' in PASS is realized at the system-architecture level. The system adopts a positioning-driven communication mechanism, where the positioning results are utilized to reconfigure the PA deployment.

The main theoretical contributions related to the proposed positioning approach in this paper are outlined below:
\begin{itemize}
  \item We employ PASS to enable uplink transmission, where ranging is performed by utilizing the received signal strength indicator (RSSI) technique. Unlike traditional fixed antenna systems, PAs are dynamically deployed along the waveguide, introducing a unique hybrid channel model that incorporates both free space propagation and waveguide loss. To ensure a comprehensive and systematic evaluation, the MWSP and MWMP scenarios are investigated under the assumptions of LoS channels and a single, stationary user.       
  \item For the MWSP scenario, we propose an RSSI-based ranging method and design an MWSP-based least squares (LS) positioning algorithm. To gain deeper insights, we carry out a comprehensive error analysis of the proposed MWSP-based LS positioning algorithm by providing a rigorous algebraic proof. Our analysis establishes a deterministic mathematical foundation for dynamic positioning, demonstrating that: i) The positioning performance improves when the area enclosed by PAs is larger. ii) Even if the geometric dilution of precision (GDoP) is large, the regions where the distances to three PAs are nearly equal exhibit higher positioning accuracy.
  \item For the MWMP scenario, to characterize the physical signal superposition from multiple PAs on the same waveguide, we derive the closed-form expression of the superposed signal received at AP, and we design a grid-search positioning algorithm based on the strength of the superposed signal. Furthermore, we analyze the errors of the proposed MWMP-based grid search positioning algorithm. Our analysis illustrates that: i) Increasing the number of non-parallel waveguides enhances positioning performance. ii) Positioning accuracy is further improved when fewer PAs are deployed on each waveguide.
  \item We reposition the PAs near the estimated user's location to enable downlink communication. Furthermore, we analyze the data rates under both MWSP and MWMP scenarios. Our analysis shows that the noise simultaneously degrades positioning accuracy and data rate, and the data rate will be further decreased when the positioning accuracy deteriorates, resulting in a double-impact of noise on data rate. 
  \item The simulation results confirm our analysis, and the results also demonstrate that: i) The proposed multi-waveguide PASS indoor positioning algorithms are simple but effective, which can achieve sub-meter positioning accuracy. ii) In the MWSP scenario, uniformly distributing the PAs in space improves the positioning accuracy and robustness.  iii) The single-waveguide multi-PA (SWMP) deployment fails to achieve effective positioning. iv) Due to positioning errors, the relocated PAs may deviate from the theoretical optimal positions and in the suboptimal positions. However, the proposed approaches achieve lower path loss and improved the communication performance compared with fixed antenna deployments.
\end{itemize}

\subsection{Organization and Notations}

The structure of this paper is as follows. Section \uppercase\expandafter{\romannumeral1} introduces the background, motivation and contribution of this work. Section \uppercase\expandafter{\romannumeral2} and Section \uppercase\expandafter{\romannumeral3} propose the system models, positioning algorithms and analysis results for the MWSP and MWMP scenarios, respectively. In Section \uppercase\expandafter{\romannumeral4}, we discuss the numerical evaluations of the proposed positioning algorithms. Finally, Section \uppercase\expandafter{\romannumeral5} summarizes the main findings and highlights avenues for future investigation. Throughout the paper, the notation $\left\| \cdot \right\|$ and $\left| {\cdot} \right|$ denote the modulus of vector and the absolute value operator, respectively. The superscript ${\left( {} \right)^ * }$ denotes the complex conjugate. ${\sigma ^2}$ represents the variance of the noise.

\section{MWSP System Model And Positioning Approach}

This section focuses on the system model and the proposed positioning approach in the MWSP scenario. We first present a detailed description of the system architecture. Then, we propose the positioning approach and analyze the data rate.  

\subsection{MWSP System Model}

\subsubsection{Antenna Model}

\begin{figure}[t!]
\centering
\includegraphics[width =3.2in]{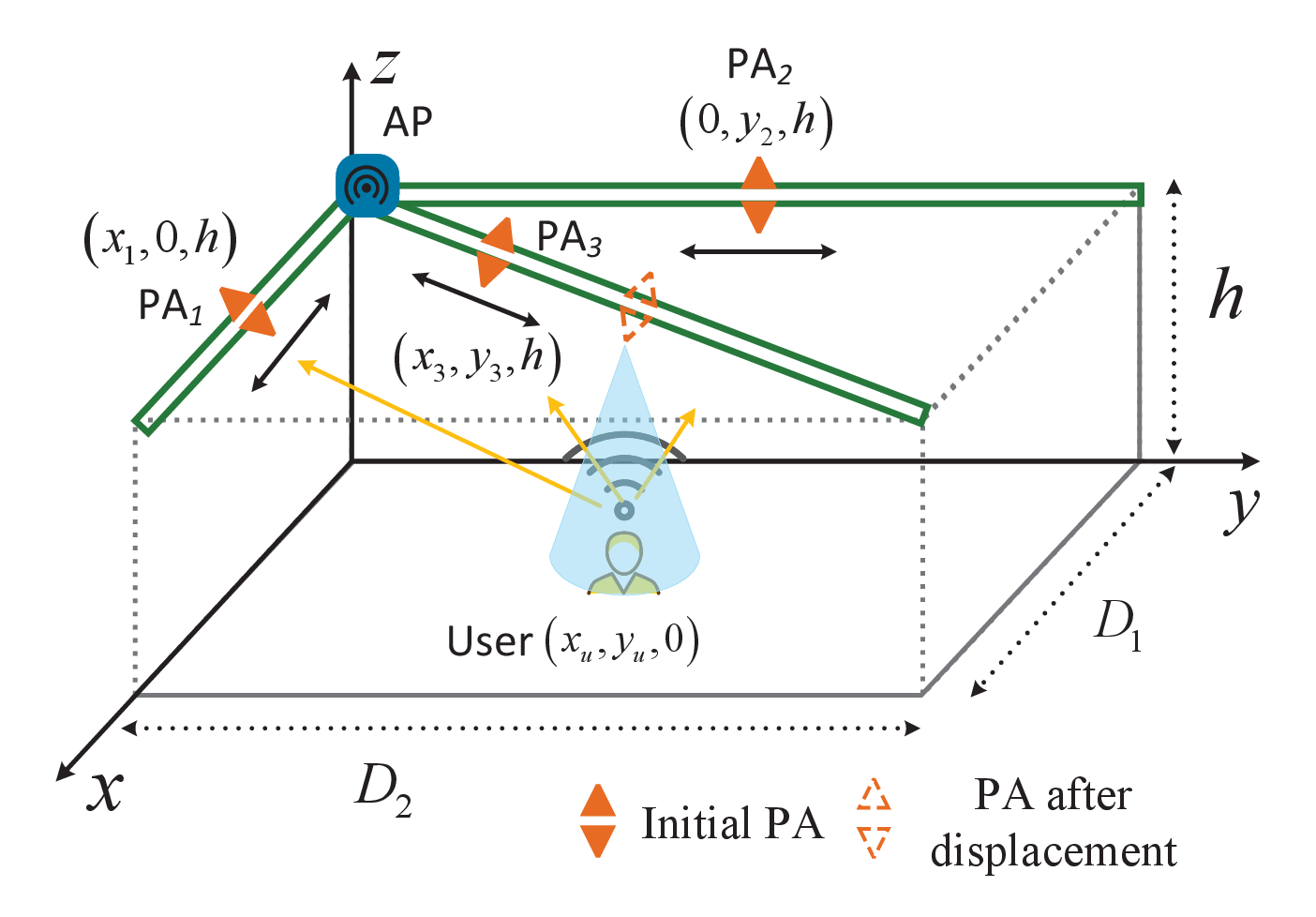}
\caption{MWSP system model.}
\label{MWSP system model}
\end{figure}

As depicted in Fig.~\ref{MWSP system model}, the room is characterized by dimensions $x \in [0, D_1]$, $y \in [0, D_2]$, and $z \in [0, h]$, with an AP located at the ceiling corner. Three waveguides are connected to the AP, extending along the ceiling edges of x-axis, the ceiling edges of y-axis, and the diagonal of the ceiling, respectively. The positions of three waveguides are defined as ${W_1} \in \left( {\left( {0,{D_1}} \right),0,h} \right)$, ${W_2} \in \left( {0,\left( {0,{D_2}} \right),h} \right)$ and ${W_3} \in \left( {\left( {0,{D_1}} \right),\left( {0,{D_2}} \right),h} \right)$, respectively. Each waveguide hosts one PA, with the position of the $k$-th PA denoted as ${L_k} = \left( {x_k,y_k,h} \right)$ for $k = 1,2,3 $, which is considered to be known. The position of the user is defined as $\left( {{x_u},{y_u},0} \right)$, which is the coordinate to be solved.

\subsubsection{Positioning Model}

Since the PAs are distributed on the ceiling, the large-scale channel is initially characterized under a LoS assumption, which aligns with the channel models adopted in prior PASS studies~\cite{PASS_2},~\cite{PASS_3}. According to the free space propagation theorem, the large-scale channel can be written as:
\begin{equation}\label{the large-scale channel in MWSP}
{h_{ls}}\left( {d_{uk}} \right) = \frac{c}{{4\pi {f_c}d_{uk}}},
\end{equation}
where $c$ represents the speed of light and $f_c$ is the carrier frequency. ${d_{uk}}$ represents the distance between the user and PA of the $k$-th waveguide, which can be written as:
\begin{equation}\label{the distance between the user and PA}
d_{uk} = \sqrt {\left( {x_k - {x_u}} \right) + {{\left( {y_k - {y_u}} \right)}^2} + {h^2}}.
\end{equation}

For the small-scale fading component, the channel coefficients are normalized, with analysis focused solely on the phase shift resulting from free-space transmission. The small-scale channel is modelled as:
\begin{equation}\label{The small-scale channel in MWSP}
{h_{ss}}\left( {d_{uk}} \right) = \exp \left( { - j\frac{{2\pi }}{\lambda }d_{uk}} \right),
\end{equation}
where $\lambda$ denotes the wavelength.

\begin{figure}[t!]
\centering
\includegraphics[width =2.8in]{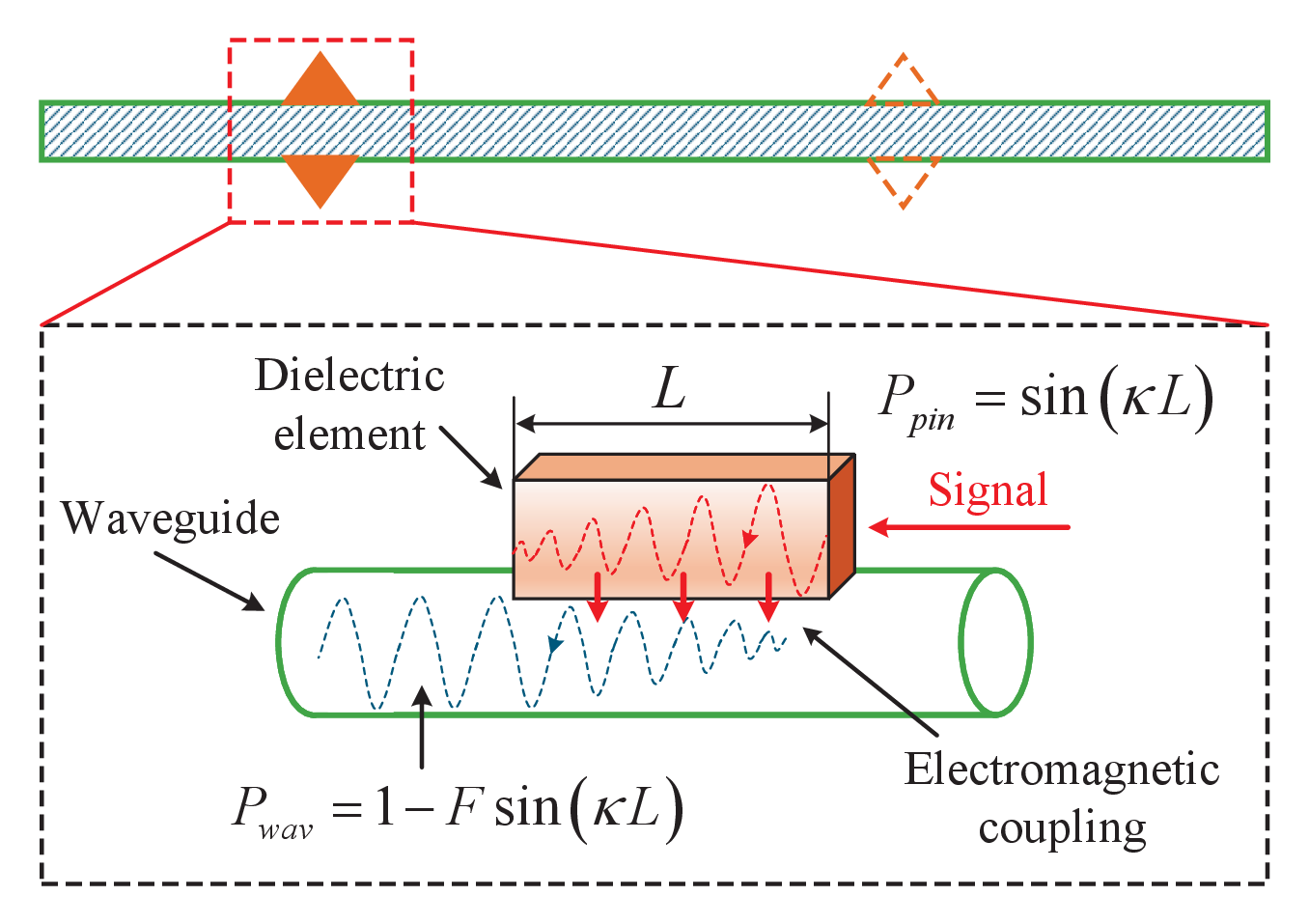}
\caption{Physical principle of PASS.}
\label{PASS}
\end{figure}

Upon reaching the PA, the signal is coupled into the waveguide. To accurately characterize the process of electromagnetic coupling, we model the PA as an open-ended dielectric element based on coupled-mode theory. As illustrated in Fig.~\ref{PASS}, the efficiency of power transfer from the dielectric element into the waveguide is determined by the physical dimensions and material properties. Let $P_{pin}$ and $P_{wav}$ represent the normalized power of the pinched dielectric element and the waveguide, respectively. The relationship between $P_{pin}$ and $P_{wav}$ can be respectively expressed as~\cite{EIM}:
\begin{equation}\label{EIM}
{P_{pin}} = \sin \left( {\kappa L} \right),
\end{equation}
and
\begin{equation}\label{EIM2}
{P_{wav}} = 1 - F\sin \left( {\kappa L} \right),
\end{equation}
where $\kappa$ and $L$ denote the coupling coefficient and coupling length of the dielectric element, respectively, and $F \le 1$ is the maximum coupling efficiency. From a physical perspective, matching the effective refractive indices of the waveguide and the dielectric element maximizes the coupling efficiency to unity. Then, the signal power in the dielectric element can be fully coupled into the waveguide by optimizing the coupling length to $\pi /\left( {2\kappa } \right)$~\cite{PASS_3}. Therefore, in this paper, we consider the ideal scenario where the coupling efficiency is 1.

When the signal propagates in the waveguide, we use the complex propagation constant to characterize the phase shift and amplitude attenuation per unit length, which is given by~\cite{Microwave_engineering}:
\begin{equation}\label{the complex propagation constant}
\gamma  = \alpha  + j\beta,
\end{equation}
where $\alpha$ denotes the attenuation constant accounting for dielectric and conductor losses, and $\beta$ is the phase constant determined by the waveguide geometry and material properties. Specifically, $\alpha$ and $\beta$ can be respectively written as:
\begin{equation}\label{Specifical constant a}
\alpha = \frac{{\pi \sqrt {{\varepsilon _r}} \tan \delta }}{\lambda },
\end{equation}
and
\begin{equation}\label{Specifical constant b}
\beta = \frac{{2\pi \sqrt {{\varepsilon _r}} }}{\lambda },
\end{equation}
where ${{\varepsilon _r}}$ and ${\tan \delta }$ denote relative permittivity and loss angle tangent, respectively.

Thus, the channel model of the $k$-th waveguide can be written as:
\begin{equation}\label{the waveguide channel in MWSP}
\begin{aligned}
{h_w}\left( {{x_k},{y_k}} \right) = \exp \left( { - \frac{{\pi \sqrt {{\varepsilon _r}} \tan \delta }}{\lambda }\left( {\sqrt {x_k^2 + y_k^2}  } \right)} \right)\\
 \times \exp \left( { - j\frac{{2\pi \sqrt {{\varepsilon _r}} }}{\lambda }\left( {\sqrt {x_k^2 + y_k^2}  } \right)} \right).
\end{aligned}
\end{equation}

The signal undergoes free-space propagation, couples into the waveguide through the PA, and is subsequently guided to the AP. The signal received from the $k$-th waveguide is given by:
\begin{equation}\label{signal received from the waveguide in MWSP}
\begin{aligned}
{r_k} &= \sqrt {{P_s}} {h_{ls}}\left( {{d_{uk}}} \right){h_{ss}}\left( {{d_{uk}}} \right){h_w}\left( {{x_k},{y_k}} \right)s + {\eta _k}\\
 &= \frac{{\sqrt {{P_s}} c}}{{4\pi {f_c}{d_{uk}}}}\exp \left( { - \frac{{\pi \sqrt {{\varepsilon _r}} \tan \delta }}{\lambda }\left( {\sqrt {x_k^2 + y_k^2} } \right)} \right)\\
 &\times \exp \left( { - j\frac{{2\pi }}{\lambda }\left( {{d_{uk}} + \sqrt {{\varepsilon _r}} \left( {\sqrt {x_k^2 + y_k^2} } \right)} \right)} \right) s +  {\eta _k},
\end{aligned}
\end{equation}
where $s$ is the narrowband single-carrier uplink signal with constant frequency and transmit power, which remains time-invariant during the measurement period. $P_s$ represents the transmission power of the signal, and $\eta_k$ is the additive white Gaussian noise (AWGN) with mean zero and variance ${\sigma ^2}$. 

\subsubsection{Communication Model}

Upon completing the uplink positioning process described in the previous subsection, we now proceed to the downlink communication phase. As illustrated in Fig.~\ref{MWSP system model}, the PA located on the waveguide closest to the user is repositioned near the user. Assuming that the estimated user position is $\left[ {{x_e},{y_e},0} \right]$, we then calculate the perpendicular distance from the estimated user position to each waveguide. We begin by projecting the estimated user position to the ceiling plane. Subsequently, we establish a two-dimensional (2D) rectangular coordinate system on the ceiling. According to the point-to-line distance theorem and the Pythagorean theorem~\cite{distance},~\cite{gougu}, the spatial distance from the estimated user to the waveguide along $x$-axis, $y$-axis and diagonal can be respectively written as:
\begin{equation}\label{dx}
\begin{aligned}
{d_x} = \sqrt {y_e^2 + {h^2}},
\end{aligned}
\end{equation}

\begin{equation}\label{dy}
\begin{aligned}
{d_y} = \sqrt {x_e^2 + {h^2}},
\end{aligned}
\end{equation}
and
\begin{equation}\label{dm}
\begin{aligned}
{d_m} = \sqrt {\frac{{{{\left( {{D_2}{x_e} - {D_1}{y_e}} \right)}^2}}}{{D_1^2 + D_2^2}} + {h^2}}.
\end{aligned}
\end{equation}

Next, we choose the waveguide with the smallest perpendicular distance to serve the user, and the PA is relocated. We define the position of the relocated PA as $\left( {x_k^{'},y_k^{'},h} \right)$. If the waveguide along the $x$-axis, $y$-axis or diagonal is selected, the relocated PA is obtained by projecting the estimated user position onto the corresponding axis, which can be written as:
\begin{equation}\label{relocated PA}
\left( {x_k^{'},y_k^{'},h} \right) = \left\{ {\begin{array}{*{20}{c}}
{\left( {{x_e},0,h} \right)},~~x-\rm{axis~waveguide},\\
{\left( {0,{y_e},h} \right)},~~y-\rm{axis~waveguide},\\
{\left( {{D_1}\xi ,{D_2}\xi ,h} \right)},~\rm{diagonal~waveguide},
\end{array}} \right.
\end{equation}
where $\xi  = \frac{{\left( {{D_1}{x_e} + {D_2}{y_e}} \right)}}{{D_1^2 + D_2^2}}$.

For the downlink communication channel, we focus primarily on free-space propagation. Since the downlink free-space channel is symmetric to the uplink channel defined in \eqref{the large-scale channel in MWSP} and \eqref{The small-scale channel in MWSP}, the received signal at user through the LoS channel is given by:
\begin{equation}\label{downlink received signal}
\begin{aligned}
{r_{u,sp}} & = \sqrt {{P_t}} h_{ls}\left(d_{uk}^{'}\right) h_{ss}\left(d_{uk}^{'}\right) {s_t} \\
&= \frac{{c\sqrt {{P_t}} }}{{4\pi {f_c}d_{uk}^{'}}}\exp \left( { - j\frac{{2\pi }}{\lambda }d_{uk}^{'}} \right){s_t},
\end{aligned}
\end{equation}
where $P_t$ and ${s_t}$ denote the transmitted power at relocated PA and communication signal, respectively. $d_{uk}^{'}$ represents the distance from the relocated PA to the user, which is given by:
\begin{equation}\label{distance between relocated PA and user}
d_{uk}^{'} = \sqrt {\left( {x_k^{'} - {x_u}} \right) + {{\left( {y_k^{'} - {y_u}} \right)}^2} + {h^2}}.
\end{equation}

\subsection{Positioning Approach and Error Analysis}

In the MWSP scenario, each waveguide is equipped with only a single PA, whereas in the MWMP scenario, the signals received by multiple PAs on the same waveguide are superimposed, both of which prevent reliable angle-of-arrival estimation. Additionally, the time-of-arrival techniques require ultra-wideband signals and specialized hardware with precise time synchronization, which is beyond the scope of our study. Consequently, as a robust, cost-effective, and widely accessible ranging method, RSSI is employed to estimate the distance between each PA and the user.

For the $k$-th waveguide, the received power of AP can be written as:
\begin{equation}\label{received power of AP in MWSP}
\begin{aligned}
{P_{rk}} &= {r_k}r_k^* = {P_s}{\left( {\frac{{c{e^{ - \alpha \sqrt {x_k^2 + y_k^2} }}}}{{4\pi {f_c}{d_{uk}}}}} \right)^2} + {\eta _k}\eta _k^*\\
 &= {P_s}{\left( {\frac{{c{e^{ - \alpha \sqrt {x_k^2 + y_k^2} }}}}{{4\pi {f_c}{d_{uk}}}}} \right)^2} + {\zeta _k},
\end{aligned}
\end{equation}
where $\zeta _k \sim \mathcal{N}\left( {0,{\sigma ^2}} \right)$ denotes the power of AWGN.

The noisy estimated distance from the $i$-th PA to the user is expressed as:
\begin{equation}\label{MWSP-estimated distance}
{\hat d_{uk}} = \frac{{c{e^{ - \alpha \sqrt {x_k^2 + y_k^2} }}}}{{\sqrt {\frac{{{P_{rk}}}}{{{P_s}}}} 4\pi {f_c}}}.
\end{equation}

By combining (\ref{the distance between the user and PA}) and (\ref{MWSP-estimated distance}), the corresponding distance measurement equation is then derived as follows:
\begin{equation}\label{multi-position equation}
{\left( {{x_k} - {x_{u}}} \right)^2} + {\left( {{y_k} - {y_{u}}} \right)^2} + {h^2} = \frac{{{P_s}}}{{{P_{rk}}}}{\left( {\frac{{c{e^{ - \alpha \sqrt {x_k^2 + y_k^2} }}}}{{4\pi {f_c}}}} \right)^2}.
\end{equation}

According to \eqref{multi-position equation}, the distance equation in MWSP scenario can be written as: 
\begin{subequations}\label{distance equation in multi-waveguide scenarios}
\begin{align}
{{{\left( {{x_1} - {x_{u}}} \right)}^2} + {{\left( {{y_1} - {y_{u}}} \right)}^2} + {h^2} = \hat d_{u1}^2},\label{11}\\
{{{\left( {{x_2} - {x_{u}}} \right)}^2} + {{\left( {{y_2} - {y_{u}}} \right)}^2} + {h^2} = \hat d_{u2}^2},\label{22}\\
{{{\left( {{x_3} - {x_{u}}} \right)}^2} + {{\left( {{y_3} - {y_{u}}} \right)}^2} + {h^2} = \hat d_{u3}^2}.\label{33}
\end{align}
\end{subequations}

\begin{algorithm}[!ht]
    \renewcommand{\algorithmicrequire}{\textbf{Input:}}
	\renewcommand{\algorithmicensure}{\textbf{Output:}}
	\caption{MWSP-based LS Algorithm}
    \label{multi-waveguide}
    \begin{algorithmic}[1] 
        \REQUIRE  The estimated distance $\hat{d}_{uk}$ and the coordinate of the $k$-th PAs $\left( {{x_k},{y_k},h} \right)$. 
	    \ENSURE The coordinate of user $\left( {{\hat{x}_{u}},{\hat{y}_{u}},0} \right)$. 
        
        \STATE Define ${e_1} = \hat d_{u1}^2 - \hat d_{u2}^2 + \left( {x_2^2 - x_1^2} \right) + \left( {y_2^2 - y_1^2} \right)$, ${e_2} = \hat d_{u1}^2 - \hat d_{u3}^2 + \left( {x_3^2 - x_1^2} \right) + \left( {y_3^2 - y_1^2} \right)$.
        \STATE According to \eqref{linearized distance equation in multi-waveguide scenarios}, construct the matrices ${\bf{A}}$, ${\bf{X}}$, ${\bf{b}}$.  
        \STATE  The solution ${\bf{X}} = {\left( {{{\bf{A}}^T}{\bf{A}}} \right)^{ - 1}}{{\bf{A}}^T}{\bf{b}}$.
        \STATE \textbf{return} $\hat{x}_{u}$, $\hat{y}_{u}$.
    \end{algorithmic}
\end{algorithm}

Clearly, there are two unknown parameters in \eqref{distance equation in multi-waveguide scenarios}, and since the AP is able to detect signals from three waveguides, a unique feasible solution can be obtained. Algorithm~\ref{multi-waveguide} outlines the procedure of position estimation. We subtract the first equation \eqref{11} from the subsequent ones \eqref{22} and \eqref{33}, and then \eqref{distance equation in multi-waveguide scenarios} is linearized for further analysis, which is given by:   
\begin{equation}\label{linearized distance equation in multi-waveguide scenarios}
\left\{ {\begin{array}{*{20}{c}}
{2{x_{u}}\left( {{x_2} - {x_1}} \right) + 2{y_{u}}\left( {{y_2} - {y_1}} \right) = {e_1}},\\
{2{x_{u}}\left( {{x_3} - {x_1}} \right) + 2{y_{u}}\left( {{y_3} - {y_1}} \right) = {e_2}},
\end{array}} \right.
\end{equation}
where 
\begin{equation}\label{e1}
{e_1} = \hat d_{u1}^2 - \hat d_{u2}^2 + \left( {x_2^2 - x_1^2} \right) + \left( {y_2^2 - y_1^2} \right),
\end{equation}
and 
\begin{equation}\label{e2}
{e_2} = \hat d_{u1}^2 - \hat d_{u3}^2 + \left( {x_3^2 - x_1^2} \right) + \left( {y_3^2 - y_1^2} \right).
\end{equation}

From a matrix perspective, \eqref{linearized distance equation in multi-waveguide scenarios} can be written as: 
\begin{equation}\label{matrix perspective-multi}
{\bf{AX}} = {\bf{b}},
\end{equation}
where 
\begin{equation}\label{A}
{\bf{A}} = \left[ {\begin{array}{*{20}{c}}
{2\left( {{x_2} - {x_1}} \right)}&{2\left( {{y_2} - {y_1}} \right)}\\
{2\left( {{x_3} - {x_1}} \right)}&{2\left( {{y_3} - {x_1}} \right)}
\end{array}} \right],
\end{equation}
\begin{equation}\label{X}
{\bf{X}} = {\left[ {\begin{array}{*{20}{c}}
{{x_{u}}}&{{y_{u}}}
\end{array}} \right]^T},\\
\end{equation}
\begin{equation}
{\bf{b}} = {\left[ {\begin{array}{*{20}{c}}
{{e_1}}&{{e_2}}
\end{array}} \right]^T}.
\end{equation}

The multi-waveguide-based LS method seeks to obtain one single solution ${\bf{X}} = {\left[ {\begin{array}{*{20}{c}}{{{\hat x}_{u}}}&{{{\hat y}_{u}}}\end{array}} \right]^T}$, such that $\left\| {{\bf{b}} - {\bf{\hat b}}} \right\|$ is minimum. The position estimation ${\bf{X}}$ is given by:
\begin{equation}\label{position estimation-multi}
{\bf{X}} = {\left( {{{\bf{A}}^T}{\bf{A}}} \right)^{ - 1}}{{\bf{A}}^T}{\bf{b}}.
\end{equation}

We then perform a detailed error analysis of the proposed positioning algorithms. The coordinates of the user are calculated by \eqref{position estimation-multi}, and since ${{\bf{A}}}$ is a square matrix with a non-zero determinant, \eqref{position estimation-multi} can be rewritten as:
\begin{equation}\label{rewritten position estimation}
\begin{aligned}
{\bf{X}} = {{\bf{A}}^{ - 1}}{\bf{b}}.
\end{aligned}
\end{equation}

We define the observation vector ${\bf{b}}$ as:
\begin{equation}\label{define the observation vector}
\begin{aligned}
{\bf{b}} = {\bf{A}}{{\bf{X}}_t} + {\bf{g}},
\end{aligned}
\end{equation}
where ${{\bf{X}}_t}$ represents the error-free solution vector, and ${\bf{g}}$ denotes error vector.

Then, \eqref{position estimation-multi} can be transformed into:
\begin{equation}\label{transformed position estimation}
\begin{aligned}
{\bf{X}} = {{\bf{A}}^{ - 1}}\left( {{\bf{A}}{{\bf{X}}_t} + {\bf{g}}} \right) = {{\bf{X}}_t} + {{\bf{A}}^{ - 1}}{\bf{g}}.
\end{aligned}
\end{equation}

Therefore, the positioning errors are caused by the error amplification factor ${{\bf{A}}^{ - 1}}$ and the error in the observation ${\bf{b}}$.

\textbf{i) Errors caused by matrix ${{\bf{A}}^{ - 1}}$}: According to~\cite{matrix}, the inverse of matrix ${{\bf{A}}}$ can be written as:
\begin{equation}\label{inverse of A}
\begin{aligned}
{{\bf{A}}^{ - 1}} = \frac{1}{{\det \left( {\bf{A}} \right)}}{{\bf{A}}^ * },
\end{aligned}
\end{equation}
where ${\det \left( {\bf{A}} \right)}$ and ${{\bf{A}}^ * }$ denote the determinant and adjoint matrix of ${\bf{A}}$, respectively.

Thus, the matrix ${{\bf{A}}^{ - 1}}$ is inversely proportional to the determinant. According to the expression in \eqref{A}, the determinant of matrix ${\bf{A}}$ can be written as:
\begin{equation}\label{determinant of A}
\begin{aligned}
\det \left( {\bf{A}} \right) = 4\left[ {\left( {{x_2} - {x_1}} \right)\left( {{y_3} - {y_1}} \right) - \left( {{x_3} - {x_1}} \right)\left( {{y_2} - {y_1}} \right)} \right].
\end{aligned}
\end{equation}

\begin{lemma}\label{lemma1:determinant of A and the area}
We first define the triangle enclosed by the spatial coordinates $\left( {{x_1},{y_1}} \right)$, $\left( {{x_2},{y_2}} \right)$ and $\left( {{x_3},{y_3}} \right)$ as ${\Delta _{p}}$, and we define the area of ${\Delta _{p}}$ as ${S_{{\Delta _{p}}}}$. We can then derive that the determinant of ${\bf{A}}$ is eight times the area enclosed by three PAs, i.e., $\det \left( \mathbf{A} \right) = 8{S_{{\Delta _p}}}$.
\begin{proof}
Please refer to Appendix~A.
\end{proof}
\end{lemma}

\begin{remark}\label{remark error and A}
Equation \eqref{transformed position estimation} and Lemma~\ref{lemma1:determinant of A and the area} indicate that the larger the area enclosed by PAs, the smaller the error amplification matrix ${{\bf{A}}^{ - 1}}$, thereby suppressing error amplification and enabling stable, accurate positioning.
\end{remark}

\textbf{ii) Errors caused by observation vector ${\bf{b}}$}: From the definitions in \eqref{e1} and \eqref{e2}, the observation errors can be written as:
\begin{equation}\label{delta e1}
\begin{aligned}
\Delta {e_1} &= \left| {\left( {\hat d_{u1}^2 - d_{u1}^2} \right) - \left( {\hat d_{u2}^2 - d_{u2}^2} \right)} \right|\\
 &= \left| {\left( {{{\left( {{d_{u1}} + {\delta _1}} \right)}^2} - d_{u1}^2} \right) - \left( {{{\left( {{d_{u2}} + {\delta _2}} \right)}^2} - d_{u2}^2} \right)} \right|\\
 &= \left| {\left( {2{d_{u1}}{\delta _1} + \delta _1^2} \right) - \left( {2{d_{u2}}{\delta _2} + \delta _2^2} \right)} \right|\\
 &\approx \left| {2\left( {{d_{u1}}{\delta _1} - {d_{u2}}{\delta _2}} \right)} \right|,
\end{aligned}
\end{equation}
and
\begin{equation}\label{delta e2}
\begin{aligned}
\Delta {e_2} &= \left| {\left( {\hat d_{u1}^2 - d_{u1}^2} \right) - \left( {\hat d_{u3}^2 - d_{u3}^2} \right)} \right|\\
 &= \left| {\left( {{{\left( {{d_{u1}} + {\delta _1}} \right)}^2} - d_{u1}^2} \right) - \left( {{{\left( {{d_{u3}} + {\delta _3}} \right)}^2} - d_{u3}^2} \right)} \right|\\
 &= \left| {\left( {2{d_{u1}}{\delta _1} + \delta _1^2} \right) - \left( {2{d_{u3}}{\delta _3} + \delta _3^2} \right)} \right|\\
 &\approx \left| {2\left( {{d_{u1}}{\delta _1} - {d_{u3}}{\delta _3}} \right)} \right|,
\end{aligned}
\end{equation}
where ${\delta _1}$, ${\delta _2}$ and ${\delta _3}$ represent the distance measurement error, which are much smaller than the distance.  

Thus, the error vector $\bf g$ can be approximated as:
\begin{equation}\label{error vector}
\begin{aligned}
{\bf{g}} = 2{\left[ {\begin{array}{*{20}{c}}
{\left( {{d_{u1}}{\delta _1} - {d_{u2}}{\delta _2}} \right)}&{\left( {{d_{u1}}{\delta _1} - {d_{u3}}{\delta _3}} \right)}
\end{array}} \right]^T}.
\end{aligned}
\end{equation}

\begin{remark}\label{remark distance error}
According to Lemma~\ref{lemma1:determinant of A and the area} and the error vector in \eqref{error vector}, the positioning accuracy is determined by the area enclosed by the coordinates of three PAs and the distance differences between the user and three PAs.  
\end{remark}

It is important to note that while the error analysis is derived for the LS estimator, the resulting insights into the system geometry are of general significance. Unlike maximum likelihood estimators and deep learning-based solvers that provide numerical results without explaining the underlying causality, the proposed LS analysis offers an explicit algebraic structure, revealing the fundamental geometric principles intrinsic to the PASS architecture. As a result, the relative performance trends identified by our analysis provide universal guidelines for PA deployment, which are applicable regardless of the specific estimator considered.

\subsection{Data Rate}

In MWSP scenario, only one PA is activated on each waveguide. Based on the results of uplink positioning, the PA on the waveguide nearest to the user is relocated to the user’s projection on that waveguide to enable downlink communication. According to the received signal in \eqref{downlink received signal}, the received signal-to-noise ratio (SNR) in MWSP scenario can therefore be expressed as:
\begin{equation}\label{SNR in MWSP}
SN{R_{sp}} = {\left( {\frac{c}{{4\pi {f_c}d_{uk}^{'}}}} \right)^2}\frac{{{P_t}}}{{{\sigma ^2}}}.
\end{equation}

Based on the Shannon capacity theory, the achievable data rate in the MWSP scenario can be written as:
\begin{equation}\label{channel capacity in MWSP}
\begin{aligned}
{R_{sp}} = {\log _2}\left( {1 + {{\left( {\frac{c}{{4\pi {f_c}d_{uk}^{'}}}} \right)}^2}\frac{{{P_t}}}{{{\sigma ^2}}}} \right).
\end{aligned}
\end{equation}

\section{MWMP System Model And Positioning Approach}

To enhance the communication performance, this section examines the system model and positioning approach when multiple PAs are activated simultaneously in the same waveguide.

\subsection{MWMP System Model}

\subsubsection{Antenna Model}

\begin{figure}[t!]
\centering
\includegraphics[width =3.3in]{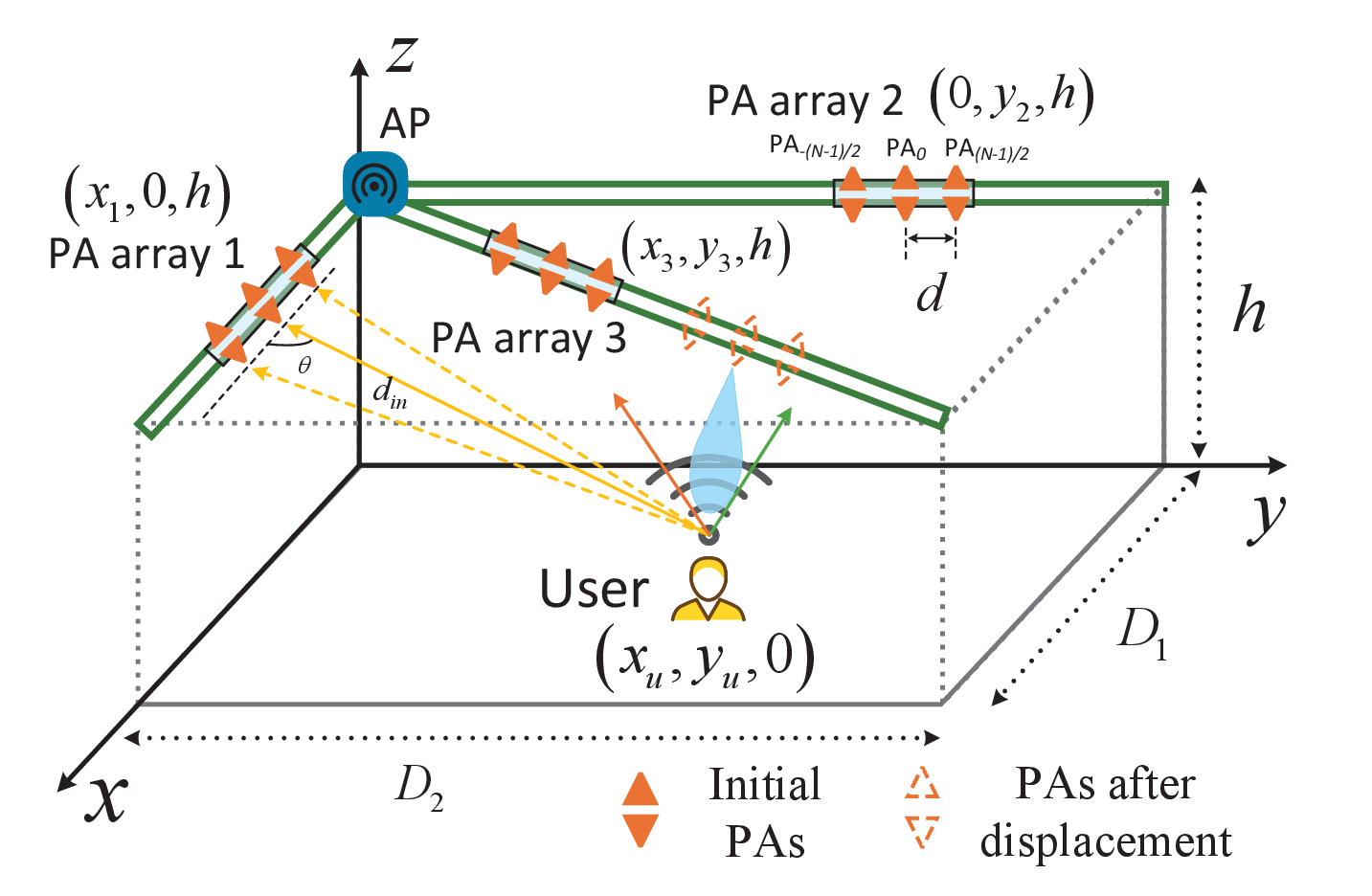}
\caption{MWMP system model.}
\label{MWMP system model}
\end{figure}

In the MWMP scenario, the spatial dimensions of the room and the waveguide distribution are identical to that of the MWSP scenario. As depicted in Fig.~\ref{MWMP system model}, the PA array with $N$ elements is located on each waveguide to serve the randomly distributed user, where the position of the user is the same as the MWSP scenario. Without loss of generality, it is assumed that $N$ is an odd integer, then the center of the array is the middle PA. The location of PA array center of the $i$-th waveguide is defined as $\left( {{x_i},{y_i},h} \right)$, and the distance $d$ between the PA elements is identical. 

\subsubsection{Positioning Model}

For far-field scenario\footnote{Utilizing the parameters in Section~IV and the definition in Section~III.B ($f_c=2.4~\mathrm{GHz}$, $\varepsilon_r=2.08$, $d=\lambda/\sqrt{\varepsilon_r}$ and the room dimensions are $6{\rm{ m}} \times 10{\rm{ m}} \times 3{\rm{ m}}$), the Fraunhofer distance can be calculated as $d_F=2(Nd)^2/\lambda=2N^2\lambda/\varepsilon_r$, yielding $d_F\approx 1.08~\mathrm{m}$ for $N=3$, and $d_F\approx 3.00~\mathrm{m}$ for $N=5$. Therefore, the user-PA distance typically exceeds the Fraunhofer distance, validating the far-field model.}, the large-scale fading from the user to each PA element in a waveguide can be approximated as identical due to the relatively small array aperture compared to the link distance. Similar to \eqref{the large-scale channel in MWSP}, the large-scale channel between the user and the $i$-th waveguide can be modeled as:
\begin{equation}\label{the large-scale channel}
{h_{ls}}\left( {{d_{ui}}} \right) = \frac{c}{{4\pi {f_c}{d_{ui}}}},
\end{equation}
where $d_{ui}$ represents the distance between the user and the array center of the $i$-th waveguide, which can be written as:
\begin{equation}\label{the distance between the user and the array center}
{d_{ui}} = \sqrt {\left( {{x_i} - {x_u}} \right) + {{\left( {{y_i} - {y_u}} \right)}^2} + {h^2}}.
\end{equation}

The small-scale channel can be written as:
\begin{equation}\label{The small-scale channel}
{h_{ss}}\left( {d_{in}} \right) = \exp \left( { - j\frac{{2\pi }}{\lambda }{d_{in}}} \right),
\end{equation}
where ${d_{in}}$ denotes the distance from the user to the $n$-th PA in the array of the $i$-th waveguide, which is given by:
\begin{equation}\label{the distance between the user and the PA}
d_{in} = \sqrt {d_{ui}^2 + {n^2}{d^2} - 2{d_{ui}} n d\cos {\theta _i}},
\end{equation} 
where $n =  - \frac{{N - 1}}{2},..., 0,...,\frac{{N - 1}}{2}$. ${\theta _i}$ denotes the angle between signal and PA array center of the $i$-th waveguide, with ${\theta _i} \in \left( {0,\pi } \right)$. 

The channel model of the $i$-th waveguide is given by:
\begin{equation}\label{ the waveguide channel}
\begin{aligned}
{h_w}\left( {{x_i},{y_i}} \right) = \exp \left( { - \frac{{\pi \sqrt {{\varepsilon _r}} \tan \delta }}{\lambda }\left( {\sqrt {x_i^2 + y_i^2}  + nd} \right)} \right)\\
 \times \exp \left( { - j\frac{{2\pi \sqrt {{\varepsilon _r}} }}{\lambda }\left( {\sqrt {x_i^2 + y_i^2}  + nd} \right)} \right).
\end{aligned}
\end{equation}

The signal received from the $i$-th waveguide in AP can be expressed as:
\begin{equation}\label{signal received from the waveguide}
\begin{aligned}
r_i &= \sqrt{P_s} {h_{ls}}\left( {{d_{ui}}} \right){h_{ss}}\left( {{d_{in}}} \right){h_w}\left( {{x_i},{y_i}} \right)s +  \eta_i\\
  &= \sqrt{P_s} \frac{c}{{4\pi {f_c}{d_{ui}}}}\sum\limits_{n =  - \left( {N - 1} \right)/2}^{\left( {N - 1} \right)/2} {{E_n^i}{s}} + \eta_i, 
\end{aligned}
\end{equation}
where
\begin{equation}\label{En}
\begin{aligned}
&E_n^i = \exp \left( { - \frac{{\pi \sqrt {{\varepsilon _r}} \tan \delta }}{\lambda }\left( {\sqrt {x_i^2 + y_i^2}  + nd} \right)} \right)\\
 &\times \exp \left( { - j\frac{{2\pi }}{\lambda }\left( {d_{in} + \sqrt {{\varepsilon _r}} \left( {\sqrt {x_i^2 + y_i^2}  + nd} \right)} \right)} \right).
\end{aligned}
\end{equation}

\subsubsection{Communication Model}

With the uplink positioning process in the previous subsection completed, the system proceeds to the downlink communication phase. As depicted in Fig.~\ref{MWMP system model}, the PA array on the waveguide nearest to the user is relocated close to the user. Similar to the MWSP scenario, the estimated user position is projected onto the nearest waveguide, and the array center $\left( {x_i^{'},y_i^{'},h} \right)$ is subsequently derived, with an expression identical to that in \eqref{relocated PA}. Since the location of the user has been estimated by the positioning algorithm, we can adjust the position of each PA in the array so that the downlink transmitted signals are superimposed in phase at the user end. When multiple PAs serve a single user in a LoS channel, the optimized distance from array center to the $n$-th PA is given by~\cite{PASS_uplink}: 
\begin{equation}\label{optimized position}
\begin{aligned}
{l_n} = \frac{{n\lambda \left( {2{d_0} + n\lambda } \right)}}{{2\left( {{d_0} + n\lambda } \right)}},
\end{aligned}
\end{equation}
where $d_0$ denotes the distance of the user and the nearest waveguide.

Subsequently, the coordinate of the $n$-th PA can be given by:
\begin{equation}\label{the coordinate of the n-th PA}
\left( {{x_n},{y_n},h} \right) = \left\{ {\begin{array}{*{20}{c}}
{\left( {{x_e} + {l_n},0,h} \right)},~~x-\rm{axis},\\
{\left( {0,{y_e} + {l_n},h} \right)},~~y-\rm{axis},\\
{\left( {{D_1}\left( {\xi  + {l_n}\mu } \right),{D_2}\left( {\xi  + {l_n}\mu } \right),h} \right)},~\rm{diagonal},
\end{array}} \right.
\end{equation}
where $\mu  = \frac{1}{{\sqrt {D_1^2 + D_2^2} }}$.

We assume equal transmit power at each PA, focusing primarily on free-space propagation. Then, the superimposed signal received by user can be expressed as:
\begin{equation}\label{signal received by user}
\begin{aligned}
{r_{u,mp}} &= \sum\limits_n {\sqrt {{P_t}} {h_{ls}}\left( {d_{in}^{'}} \right){h_{ss}}\left( {{l_n} + d_{in}^{'}} \right){s_t}}\\
&= \sum\limits_n {\sqrt {{P_t}} \frac{c}{{4\pi {f_c}{d_{in}^{'}}}}\exp \left( { - j\frac{{2\pi }}{\lambda }\left( {{l_n} + d_{in}^{'}} \right)} \right)} {s_t}\\
 &= \sqrt {{P_t}} {s_t}\sum\limits_n {\frac{c}{{4\pi {f_c}d_{in}^{'}}}}, 
\end{aligned}
\end{equation}
where $d_{in}^{'}$ represents the distance between the user and the $n$-th PA of relocated array, which can be written as:
\begin{equation}\label{the distance between the user and the array center}
d_{in}^{'} = \sqrt {\left( {{x_n} - {x_u}} \right) + {{\left( {{y_n} - {y_u}} \right)}^2} + {h^2}}.
\end{equation}

\subsection{Positioning Approach}

From \eqref{signal received from the waveguide} and \eqref{En}, we can observe that the power of the received signal is related the user's position. In order to simplify the analysis and obtain the closed-form expression of the superimposed received signal, we define $\sqrt {{\varepsilon _r}}d = \lambda $. Then, $d = \lambda /\sqrt {{\varepsilon _r}} $, and the distance $d$ satisfies $d \ll 1$. Therefore, the waveguide loss from each PA to the AP can be approximated as the loss from the array center to AP. Then, the superimposed received signal of the $i$-th waveguide can be written as:
\begin{equation}\label{the superimposed received signal of the waveguide along the x-axis}
\begin{aligned}
{r_i} &= \frac{{\sqrt {{P_s}} c{e^{ - \alpha \sqrt {x_i^2 + y_i^2} }}}}{{4\pi {f_c}{d_{ui}}}}\\
 &\times \sum\limits_{n =  - \left( {N - 1} \right)/2}^{\left( {N - 1} \right)/2} {\exp \left( { - j\frac{{2\pi }}{\lambda }\left( {{d_{in}} + \sqrt {{\varepsilon _r}}\sqrt {x_i^2 + y_i^2} } \right)} \right)} {s} + {\eta _i}.
\end{aligned}
\end{equation}

Expanding the distance ${d_{in}}$ in \eqref{the distance between the user and the PA} by using a first-order Taylor series, the linearized distance is derived as follows:
\begin{equation}\label{first-order Taylor series}
{d_{in}} = {d_{ui}} - nd\cos {\theta_i}.
\end{equation}

By substituting \eqref{first-order Taylor series} into \eqref{the superimposed received signal of the waveguide along the x-axis}, the superimposed received signal can be rewritten as:
\begin{equation}\label{rewritten superimposed received signal x}
\begin{aligned}
{r_i} &= \frac{{\sqrt {{P_s}} c{e^{ - \alpha \sqrt {x_i^2 + y_i^2} }}}}{{4\pi {f_c}{d_{ui}}}}\\
 &\times \exp \left( { - j\frac{{2\pi }}{\lambda }\left( {{d_{ui}} + \sqrt {{\varepsilon _r}}\sqrt {x_i^2 + y_i^2} } \right)} \right)\\
 &\times \sum\limits_{n =  - \left( {N - 1} \right)/2}^{\left( {N - 1} \right)/2} {\exp \left( {j\frac{{2\pi }}{\lambda }nd\cos {\theta _i}} \right)} {s} + {\eta _i}.
\end{aligned}
\end{equation}

\begin{lemma}\label{lemma2:cos_i}
The cosine values of angles between the signal and three waveguides can be respectively expressed as:
\begin{equation}\label{cosine of the angle-x}
\cos {\theta _1}  = \frac{{{x_u} - {x_1}}}{{{d_{u1}}}},
\end{equation}

\begin{equation}\label{cosine of the angle-y}
\begin{aligned}
\cos {\theta _2} =  \frac{{{y_u} - {y_2}}}{{{d_{u2}}}},
\end{aligned}
\end{equation}
and
\begin{equation}\label{cosine of the angle-m}
\begin{aligned}
\cos {\theta _3} =  \frac{{{D_1}\left( {{x_u} - {x_3}} \right) + {D_1}\left( {{y_u} - {y_3}} \right)}}{{{d_{u2}}\sqrt {D_1^2 + D_2^2} }}.
\end{aligned}
\end{equation}

\begin{proof}
Please refer to Appendix~B.
\end{proof}
\end{lemma}

\begin{lemma}\label{lemma3:superimposed received signal closed-form}
For the waveguide distributed on the ceiling, the closed-form expression of the superimposed received signal can be written as:
\begin{equation}\label{closed-form expression of the superimposed received signal in lemma2}
\begin{aligned}
{r_i} &= \frac{{\sqrt {{P_s}} c{e^{ - \alpha \sqrt {x_i^2 + y_i^2} }}}}{{4\pi {f_c}{d_{ui}}}}\frac{{\sin \left( {\frac{{N{\psi _i}}}{2}} \right)}}{{\sin \left( {\frac{{{\psi _i}}}{2}} \right)}}\\
 &\times \exp \left( { - j\frac{{2\pi }}{\lambda }\left( {{d_{ui}} + \sqrt {{\varepsilon _r}}\sqrt {x_i^2 + y_i^2} } \right)} \right){s} + {\eta _i},
\end{aligned}
\end{equation}
where
\begin{align*}
{\psi _i} = \frac{{2\pi d\cos {\theta _i}}}{\lambda }.
\end{align*}
\begin{proof}
Please refer to Appendix~C.
\end{proof}
\end{lemma}

According to the closed-form expression of the superimposed received signal, the power of the received signal can be derived as follows:
\begin{equation}\label{theoretical power-x}
\begin{aligned}
{P_{thi}} &= {r_i}r_i^* = {P_s} {\left( {\frac{{c{e^{ - \alpha \sqrt {x_i^2 + y_i^2} }}}}{{4\pi {f_c}{d_{ui}}}}\frac{{\sin \left( {\frac{{N{\psi _i}}}{2}} \right)}}{{\sin \left( {\frac{{{\psi _i}}}{2}} \right)}}} \right)^2}+ {\eta _i}\eta _i^*\\ 
&= {P_s} {\left( {\frac{{c{e^{ - \alpha \sqrt {x_i^2 + y_i^2} }}}}{{4\pi {f_c}{d_{ui}}}}\frac{{\sin \left( {\frac{{N{\psi _i}}}{2}} \right)}}{{\sin \left( {\frac{{{\psi _i}}}{2}} \right)}}} \right)^2} + {\zeta _i},
\end{aligned}
\end{equation}
where $\zeta _i \sim \mathcal{N}\left( {0,{\sigma ^2}} \right)$ denotes the power of AWGN.

Assuming that the measured power is ${P_{ri}}$, the superimposed power-based positioning equation can be given by:
\begin{equation}\label{positioning equation-x}
\begin{aligned}
{P_{ri}}\left( {{x_u},{y_u}} \right) = {P_s}{\left( {\frac{{c{e^{ - \alpha \sqrt {x_i^2 + y_i^2} }}}}{{4\pi {f_c}{d_{ui}}}}\frac{{\sin \left( {\frac{{N{\psi _i}}}{2}} \right)}}{{\sin \left( {\frac{{{\psi _i}}}{2}} \right)}}} \right)^2} + {\zeta _i}.
\end{aligned}
\end{equation}

\begin{figure}[htbp]
\centering
\subfigure[The waveguide along x-axis.]{\includegraphics[width =1.7in]{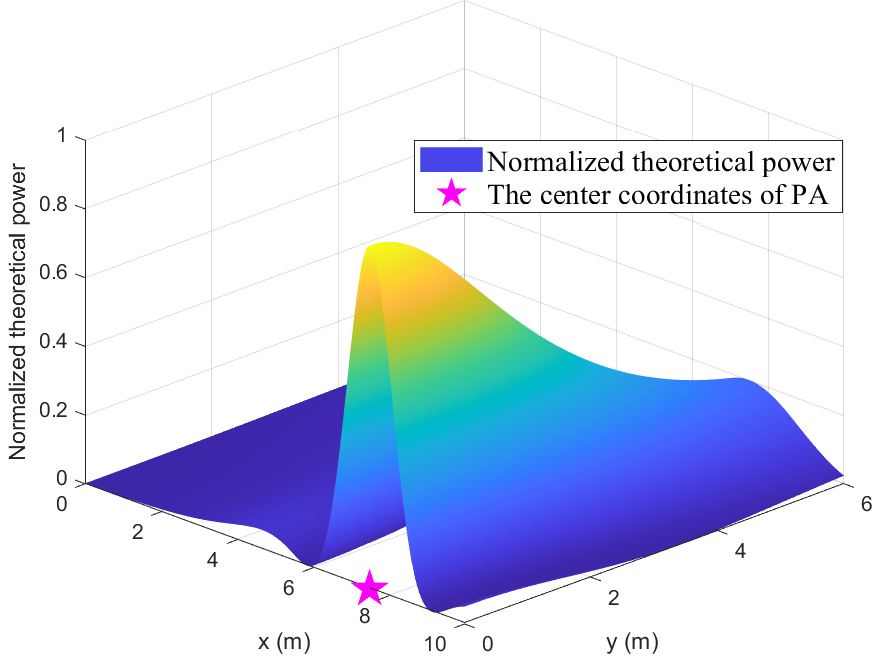}}
\subfigure[The waveguide along y-axis.]{\includegraphics[width =1.7in]{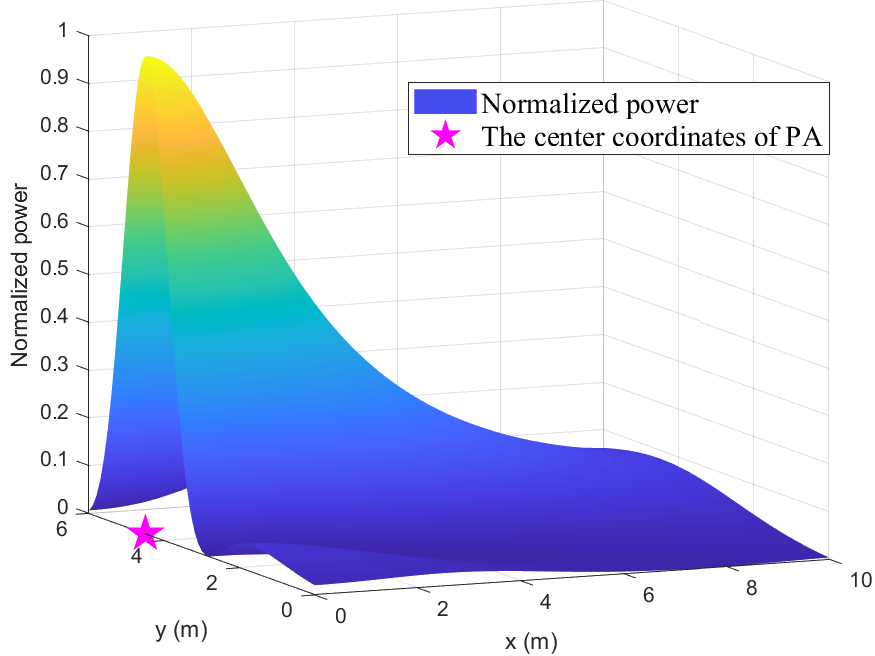}}
\subfigure[The waveguide along diagonal.]{\includegraphics[width =1.7in]{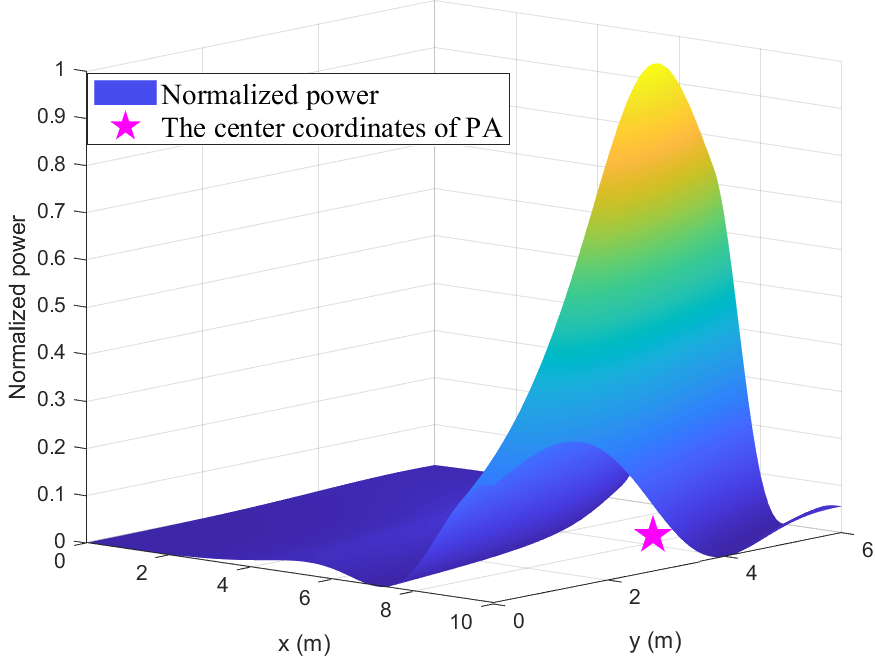}}
\caption{Schematic diagram of normalized superposition power, with $N=3$.}
\label{Normalized Power}
\end{figure}

However, since $\cos {\theta _i}$ is symmetrical in ${\theta _i} \in \left( {0,\pi } \right)$, ${\left( {\frac{{\sin \left( {\frac{{N{\psi _i}}}{2}} \right)}}{{\sin \left( {\frac{{{\psi _i}}}{2}} \right)}}} \right)^2}$ is a symmetric function. Therefore, the received power is also symmetrical, which is shown in Fig.~\ref{Normalized Power}.

\begin{remark}\label{remark single waveguide}
According to the results in \eqref{theoretical power-x}, since the theoretical power is symmetrical, a single waveguide cannot achieve high-precision positioning. High positioning accuracy is achieved only when the user is located in an area passing through the center of the PA array and perpendicular to the waveguide propagation direction.
\end{remark}

\begin{remark}\label{remark multiple parallel waveguides}
When multiple parallel waveguides are distributed on the ceiling, the positioning performance is poor because parallel waveguides share identical symmetry forms, differing only in symmetry points, making the symmetry difficult to eliminate.
\end{remark}

\begin{algorithm}[!ht]
    \renewcommand{\algorithmicrequire}{\textbf{Input:}}
	\renewcommand{\algorithmicensure}{\textbf{Output:}}
	\caption{MWMP-based Grid Search Algorithm}
    \label{Positioning Algorithm}
    \begin{algorithmic}[1] 
        \REQUIRE  The coordinate of each PA array center, the transmitted signal power and the measured received signal power of each waveguide. 
	    \ENSURE The coordinate of user $\left( {{\hat{x}_{u}},{\hat{y}_{u}},0} \right)$. 
        \STATE Define the error function as ${e_r}\left( {{x_u},{y_u}} \right) = \left| {{P_{r1}} - {P_{th1}}} \right| + \left| {{P_{r2}} - {P_{th2}}} \right| + \left| {{P_{r3}} - {P_{th3}}} \right|$. 
        \STATE Divide the room into several coarse grid points.
        \STATE Compute the received power $P_{thi}$ for all possible grid points utilizing equation derived from \eqref{theoretical power-x}.
        \STATE Evaluate the error function ${e_r}\left( {{x_u},{y_u}} \right)$ at each grid point as defined previously.
        \STATE Apply iterative refinement by subdividing the area around the grid point with minimal initial error to form a finer search grid, where ${x_u} \in \left[ {0,{D_1}} \right]$ and ${y_u} \in \left[ {0,{D_2}} \right]$.
        \STATE Repeat the iteration until the difference in the error function between two consecutive iterations falls below a preset threshold.    
        \STATE \textbf{return}: The coordinate of the grid point $\hat{x}_{u}$, $\hat{y}_{u}$.
    \end{algorithmic}
\end{algorithm}

Thus, multiple non-parallel waveguides are required to break the symmetry and enable precise positioning for user. As illustrated in Fig.~\ref{Normalized Power}, when the waveguides are not distributed in parallel, the superposition power exhibits different symmetry, which inspired us to design a multi-waveguide joint positioning algorithm. Since the expression of superimposed power is nonlinear and may exhibit multiple local optima due to the sidelobes, gradient-based methods are sensitive to initialization. Therefore, a grid-search approach is adopted to ensure convergence to the global optimum. Algorithm~\ref{Positioning Algorithm} presents the specific steps of the MWMP-based grid search algorithm.

\subsection{Positioning Error and Computational Complexity Analysis}

We further explored the errors and computational complexity of the proposed positioning algorithm. The positioning errors include waveguide loss approximation errors, far-field channel approximation errors and power symmetry point errors.

\subsubsection{Waveguide loss approximation errors}

In \eqref{the superimposed received signal of the waveguide along the x-axis}, we apply the uniform waveguide loss assumption, where the waveguide loss is approximated by using the loss from the PA array center to the AP. Since the waveguide loss is much smaller than the path loss of free space propagation, and the inter-PA distance is much smaller than unity, i.e., $d \ll 1$, the impact of waveguide loss on signal amplitude is relatively minor.  

\subsubsection{Far-field channel approximation errors}

In \eqref{first-order Taylor series}, we apply a first-order Taylor approximation for the path distance from the user to each PA, i.e, the far-field approximation. Although this approximation significantly simplifies the mathematical complexity, it introduces phase errors in the received signals. Particularly, the linearization neglects higher-order terms, leading to inaccuracies in the modeled phase shifts at each PA. Consequently, the resulting discrepancies in phase estimation translate into measurable deviations in the theoretical superimposed power.

\subsubsection{Power Symmetry Point Errors}

Although Algorithm~\ref{Positioning Algorithm} utilizes three waveguides, certain spatial configurations may still result in symmetrical power distributions, leading to ambiguous localization results. Furthermore, from \eqref{theoretical power-x}, we can observe that as the number of PAs increases, the sidelobes of the theoretical power increase. The sidelobes intensify the symmetry ambiguities and decreases positioning precision.

\begin{remark}\label{remark PAs increases}
Due to the power sidelobes, the positioning performance of the proposed algorithm decreases as the number of PAs increases. When the number of PAs is set to $ N = 1$,  the approximation coincides with the exact value.
\end{remark}

\begin{remark}\label{remark non-parallel waveguides}
Since non-parallel waveguides have different symmetry characteristics, combining non-parallel waveguides for positioning can improve accuracy. Therefore, the greater the number of non-parallel waveguides distributed on the ceiling, the better the positioning performance.
\end{remark}

To evaluate the practical feasibility of the proposed approach, we analyze the computational complexity of the MWMP-based grid search algorithm. The complexity is primarily determined by the number of grid points evaluated and the cost of computing the theoretical signal power at each point. Let $G_{coa}$ denote the number of grid points in the initial coarse search, $G_{ref}$ denote the number of points in each refinement step, and $M$ denote the number of iterations. Leveraging the closed-form expression in \eqref{theoretical power-x}, the superimposed signal power for a given waveguide can be computed by using only elementary trigonometric operations. Consequently, the computational cost per grid point is $ \mathcal O\left( 1 \right)$, which is independent of the number of PAs on each waveguide. Then, the computational complexity of Algorithm~\ref{Positioning Algorithm} can be expressed as:
\begin{equation}\label{computational complexity}
C = \mathcal O\left( {3\left( {{G_{coa}} + M{G_{ref}}} \right)} \right).
\end{equation}

This analysis demonstrates the scalability of the proposed system. Unlike conventional multi-antenna systems, where computational complexity typically increases with the number of antenna elements, our approach can deploy multiple PAs to enhance communication gain without incurring additional computational overhead in the positioning phase. Moreover, the proposed coarse-to-fine refinement achieves high-accuracy positioning with only linear growth in the number of iterations, thereby avoiding the exponential computational cost of exhaustive high-resolution grid searches.

\subsection{Data Rate}

In the MWMP scenario, multiple PAs are simultaneously activated on the waveguide to serve the user. As shown in \eqref{optimized position}, the PAs in the array are optimized to ensure that the transmitted signals can be superimposed in phase at the user. Based on the received signal in \eqref{signal received by user}, the received SNR in MWMP scenario can be written as:
\begin{equation}\label{SNR in MWMP}
SN{R_{mp}} = {\left( {\sum\limits_n {\frac{c}{{4\pi {f_c}d_{in}^{'}}}} } \right)^2}\frac{{{P_t}}}{{{\sigma ^2}}}.
\end{equation}

According to the Shannon capacity, the achievable data rate in the MWMP scenario is given by:
\begin{equation}\label{channel capacity in MWSP}
\begin{aligned}
{R_{mp}} = {\log _2}\left( {1 + {{\left( {\sum\limits_n {\frac{c}{{4\pi {f_c}d_{in}^{'}}}} } \right)}^2}\frac{{{P_t}}}{{{\sigma ^2}}}} \right).
\end{aligned}
\end{equation}

\begin{remark}\label{remark double impact}
The noise affects both positioning accuracy and data rate simultaneously. Furthermore, when the positioning accuracy deteriorates, the data rate will be more seriously affected. Therefore, the noise has a serious double-impact on data rate.
\end{remark}

\section{Simulation Results}

Here, simulations are performed to evaluate the performance of MWSP-based LS algorithm and MWMP-based grid search algorithm. The data rates in both MWSP and MWMP scenarios are also evaluated. The room dimensions are defined as $x \in [0,6]$, $y \in [0,10]$, and $z \in [0,3]$. A carrier frequency of 2.4 GHz is considered, while the user's transmit power is fixed at 0.1 W. The dielectric properties are characterized by a relative permittivity of ${\varepsilon _r = 2.08}$ and a loss tangent of $\tan \delta = 0.0004$~\cite{Microwave_engineering}. Note that since the distances between the user and different PAs or PA arrays are not the same, the received SNR differs for each link within a single measurement. Therefore, we adopt the noise power as the independent variable to evaluate the system performance.

\subsection{Positioning performance in MWSP scenario}

In the MWSP scenario, each waveguide is equipped with one PA, and three PAs are activated simultaneously. We conduct simulations to evaluate the impact of PA distribution on positioning accuracy.

\begin{figure}[t!]
\centering
\includegraphics[width =3in]{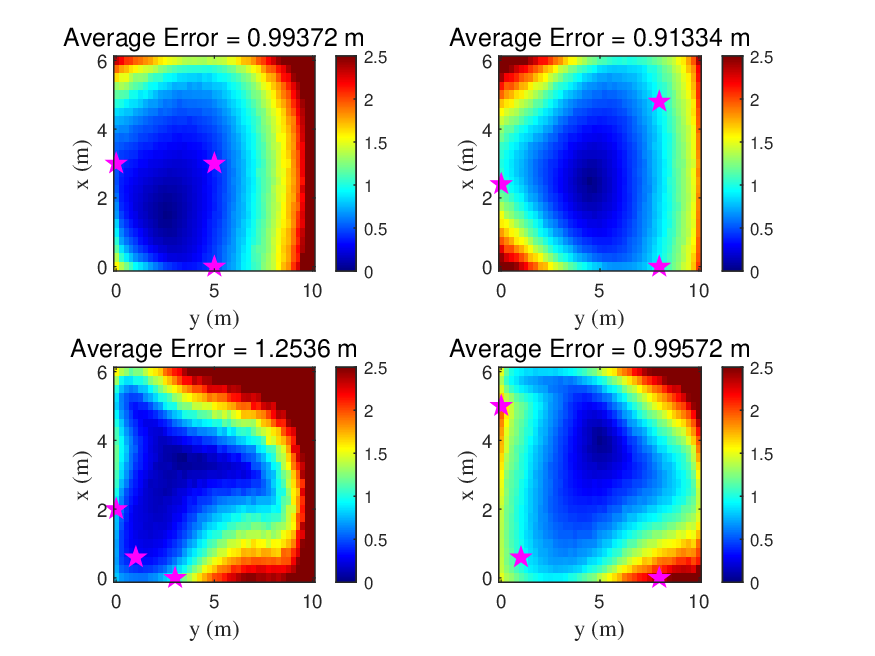}
\caption{Positioning error distribution for various PA deployment, with ${\sigma ^2}$ = -40 dBm.}
\label{multi_differPA_map}
\end{figure}

\emph{1) The impact of PAs distribution on positioning accuracy:} Fig.~\ref{multi_differPA_map} demonstrates the impact of the spatial deployment of PAs on positioning errors. The color blocks in the figure represent different positioning errors, which are measured in meters. Magenta stars represent the coordinates of PAs. When the PAs are sparsely distributed within the room, the positioning performance within the region enclosed by PAs is superior. Moreover, the larger the enclosed area, the smaller the positioning error, which verifies our~\textbf{Remark~\ref{remark error and A}}. In contrast, when the PAs are more densely distributed within the room, the region with smaller positioning error is not inside the region enclosed by PAs, but in the region with nearly equal distance to three PAs, which verifies our~\textbf{Remark~\ref{remark distance error}}. In general, a larger area enclosed by PAs results in a lower average positioning error, and the region where the distances to three PAs are nearly equal yield the best positioning performance. 

\begin{figure}[t!]
\centering
\includegraphics[width =3in]{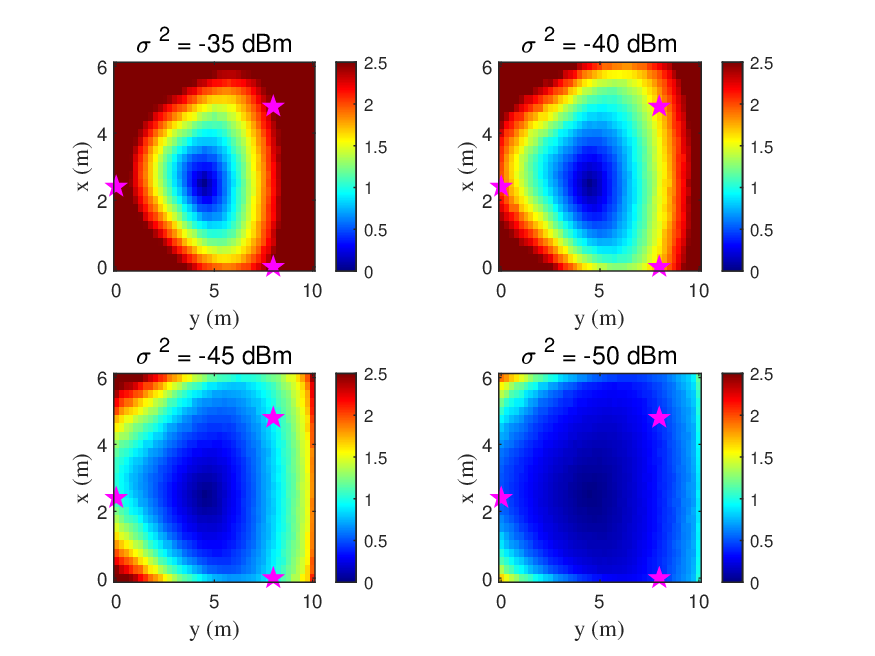}
\caption{Positioning error distribution across different noise power.}
\label{multi_differnoise_map}
\end{figure}

\emph{2) The impact of noise power on positioning accuracy:} Fig.~\ref{multi_differnoise_map} illustrates the evolution of the positioning error region as the measurement noise power ${\sigma ^2}$ is progressively reduced from –35 dBm to –50 dBm. When the noise power is high, the high-accuracy region is confined to a small vicinity around the centroid of three PAs, where the trilateration geometry is most favorable. When the noise power is reduced to ${\sigma ^2}$ = –40 dBm, the high-accuracy region expands primarily along the PA-to-PA links, which correspond to areas with minimal GDoP and thus benefit the most from noise reduction. When the noise power is further decreased to ${\sigma ^2}$ = –45 dBm, sub-meter positioning accuracy can be achieved across the region enclosed by PAs. Finally, at ${\sigma ^2}$ = –50 dBm, the high-accuracy region fills nearly the entire room. This anisotropic expansion reflects the fundamental property of multilateration: The positioning accuracy improves most rapidly in directions with low GDoP, and thus the high-accuracy region grows outward along the PA-to-PA links.

\subsection{Positioning performance in SWMP and MWMP scenarios}

In the SWMP scenario, $N$ PAs are located on the waveguide. In the MWMP scenario, $N$ PAs are located on the waveguide along $x$-axis, $y$-axis and diagonal, respectively. The PAs in SWMP and MWMP scenarios are activated simultaneously. We conduct simulations to verify the effectiveness of the proposed MWMP-based grid search algorithm.

\begin{figure}[t!]
\centering
\includegraphics[width =3in]{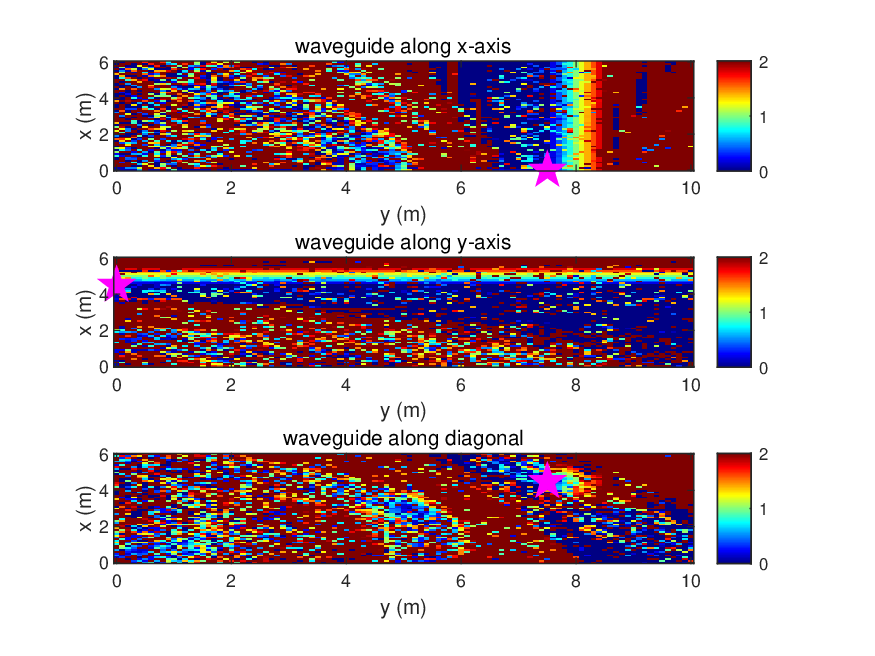}
\caption{Positioning errors in SWMP scenario, the noise power is set to ${\sigma ^2}$ = -80 dBm, with $N$ = 3.}
\label{SWMP positioning}
\end{figure}

\emph{1) Positioning errors in the SWMP scenario:} Fig.~\ref{SWMP positioning} depicts the positioning performance in the SWMP scenario. We can see that the positioning performance is poor whether the waveguide is distributed along the x-axis, y-axis or diagonal. Notably, the positioning performance is excellent when the user is located in the area centered on the PA array and oriented perpendicular to the waveguide’s propagation direction. The simulation results verify our~\textbf{Remark~\ref{remark single waveguide}}. There are countless power symmetry points on both sides of the PA array center. When the positioning algorithm infers the user's position based on the received power, the algorithm cannot distinguish these symmetry points, resulting in positioning ambiguity. 

\begin{figure}[htbp]
\centering
\subfigure[The waveguides are parallel to the y-axis.]{\includegraphics[width =3in]{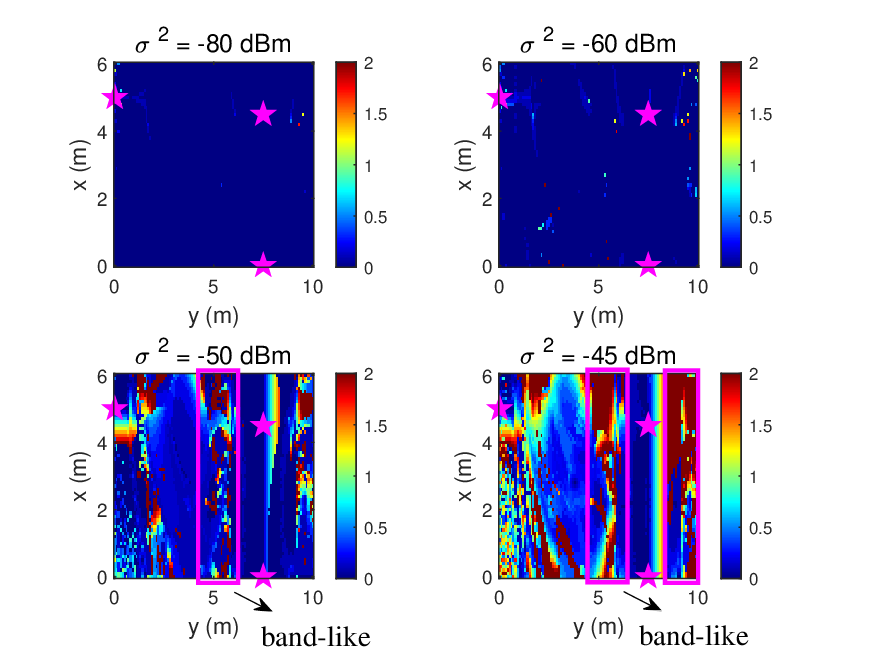} \label{fig:Onlyx}}
\subfigure[The waveguides are distributed non-parallel.]{\includegraphics[width =3in]{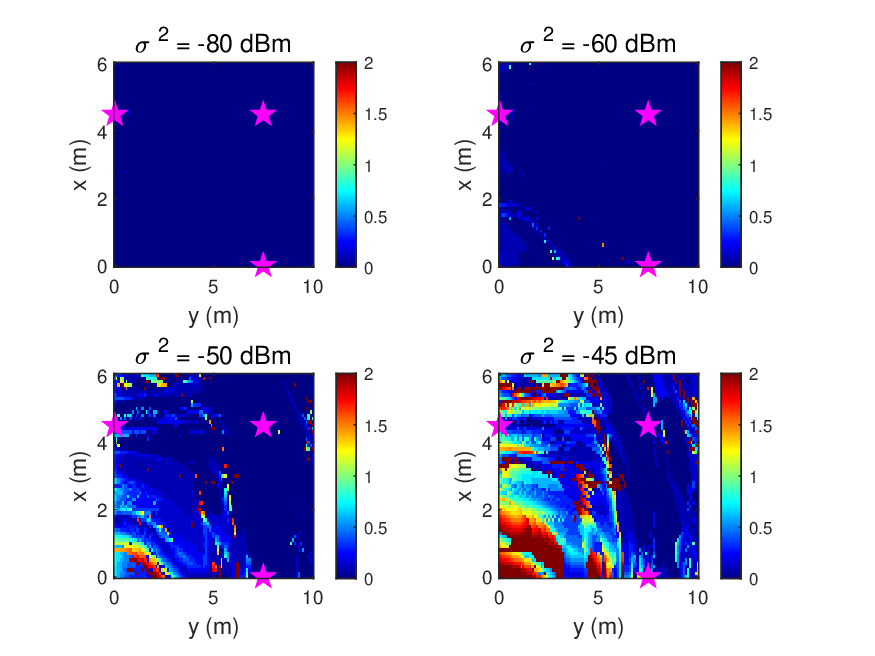} \label{fig:2DNoi}}
\caption{Positioning errors in MWMP scenario, where $N$ = 3.}
\label{MWMP positioning}
\end{figure}

\emph{2) Positioning errors in the MWMP scenario:} Fig.~\ref{MWMP positioning} demonstrates the distribution of positioning errors in the MWMP scenario with parallel and non-parallel waveguide deployments, illustrating the impact of noise power on positioning performance. When the noise power is low, the positioning errors of both deployments are low, which indicates that there is sufficient measurement information for both parallel and non-parallel waveguide deployments under high SNR conditions. However, when the noise power becomes higher, the positioning performance of parallel deployment deteriorates rapidly, while the robustness of non-parallel deployment is significantly better. As the noise increases, many band-like error zones emerge in Fig.~\ref{fig:Onlyx}, and the positioning performance is poor, which confirms our~\textbf{Remark~\ref{remark multiple parallel waveguides}}. The waveguides parallel to $y$-axis provide a strong constraint to $x$-direction but a weak constraint to $y$-direction. Thus, once the noise exceeds a certain threshold, the algorithm fails to effectively distinguish the positions of different $y$-coordinates on the same band-like regions, resulting in a systematic and widespread positioning failure. 

In contrast, as shown in Fig.~\ref{fig:2DNoi}, under the same high noise power, since the three waveguides along $x$-axis, $y$-axis and diagonal provide positioning information from different dimensions, which effectively breaks the symmetry, and thus no band-like error region are observed. The x-axis, y-axis and diagonal waveguides provide synergistic constraints on each direction, thereby making the positioning more accurate and stable, which verifies our~\textbf{Remark~\ref{remark non-parallel waveguides}}. Simulation results demonstrate that the geometric layout of waveguides is a significant factor affecting the performance and robustness of the PASS positioning system. A non-parallel, orthogonal deployment is crucial for eliminating positioning ambiguity in low SNR environments.

\begin{figure}[t!]
\centering
\includegraphics[width =3in]{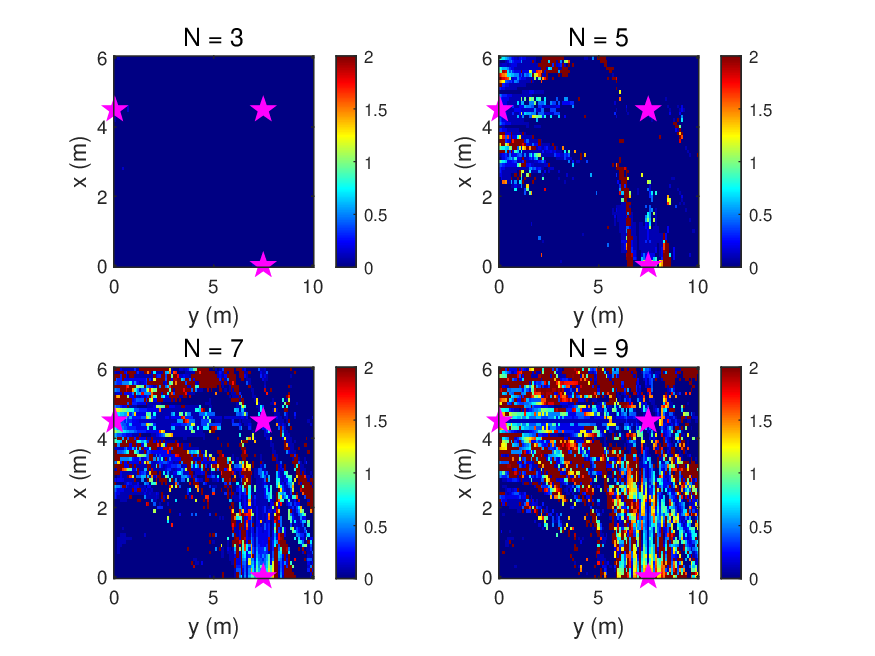}
\caption{Positioning errors under different numbers of PAs, and the noise power is set to ${\sigma ^2}$ = -80 dBm.}
\label{differNum}
\end{figure}

\emph{3) Impact of the number of PAs on positioning accuracy:} Fig.~\ref{differNum} illustrates the impact of the number of PAs on the positioning performance. We can see that as the number of PAs on each waveguide increases, the positioning performance gradually degrades. When the number of PAs is less than 5, the overall positioning error is lower than 0.5 m. When the number of PAs reaches 9, high-error area increases. This occurs because a larger number of PAs leads to more pronounced sidelobes in the theoretical power function, which increases the number of symmetry points, leading to positioning ambiguity. The simulation results verify our~\textbf{Remark~\ref{remark PAs increases}}. Notably, even when the number of PAs reaches 9, the positioning errors remain small in the strip-shaped region connecting the center of the PA array and the area in the lower left corner. These areas exhibit maximum gradient magnitudes in the theoretical power map, where the positioning sensitivity is maximized. Constrained by multiple geometric dimensions, the areas achieve significantly higher positioning accuracy, which confirms our previous research.

\begin{figure}[t!]
\centering
\includegraphics[width =3in]{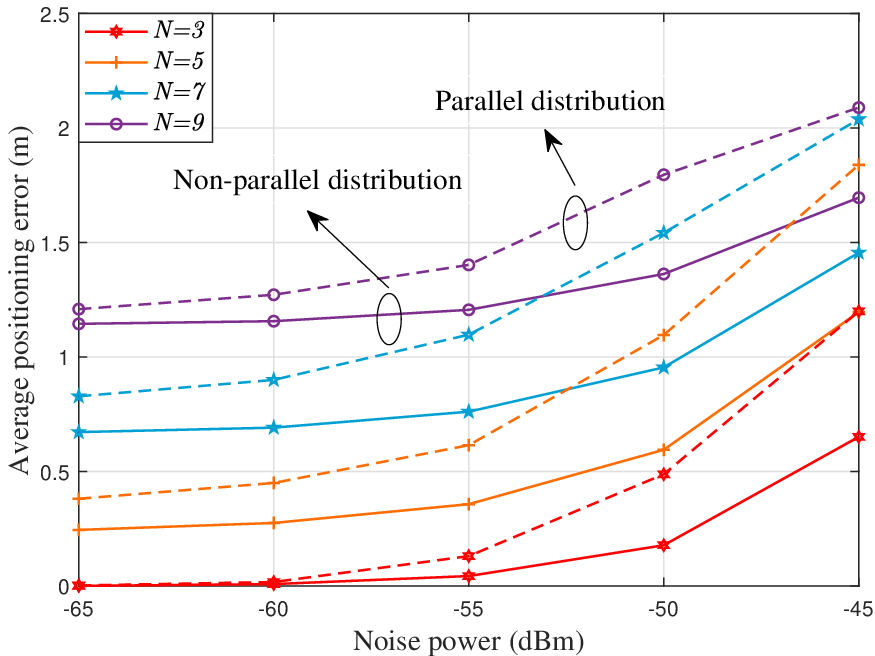}
\caption{Average positioning errors of parallel and non-parallel waveguide distributions.}
\label{ave_error}
\end{figure}

\emph{4) Comparison of positioning performance between parallel and non-parallel waveguide distribution:} As depicted in Fig.~\ref{ave_error}, the non-parallel waveguide distribution consistently outperforms the parallel distribution over the entire noise range, and the gap widens when the noise power is high, which indicates that non-parallel waveguide distribution can improve positioning accuracy and robustness. The slope of the parallel waveguide distribution is steeper than that of the non-parallel distribution, indicating that the parallel distribution is more sensitive to noise, whereas the non-parallel distribution exhibits more stable performance in high-noise regions. These simulation results also verify our~\textbf{Remark~\ref{remark non-parallel waveguides}}. In addition, as the number of PAs in the array increases, the average positioning error increases and the lower limit of positioning performance decreases, which verifies our~\textbf{Remark~\ref{remark PAs increases}}. Because the system requires an increased number of PAs in the array to enhance the beamforming degrees of freedom on the communication side, whereas a larger number of PAs leads to degraded positioning performance, there is a trade-off between positioning accuracy and communication performance.

\subsection{Data rate}

\begin{figure}[t!]
\centering
\includegraphics[width =3in]{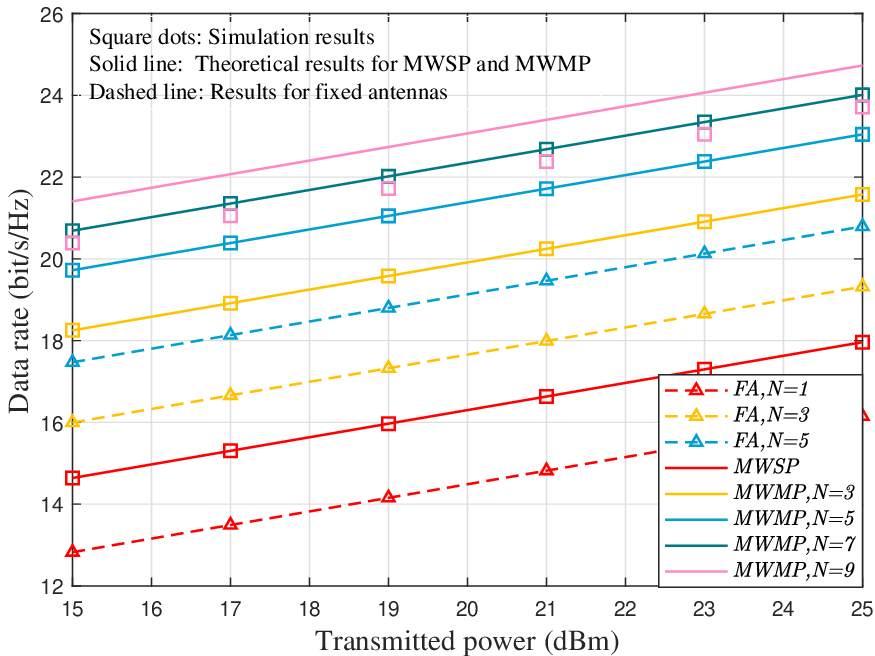}
\caption{Data rate of MWSP and MWMP scenarios versus the transmitted power, and the noise power is -80 dBm.}
\label{CvP}
\end{figure}

\emph{1) Impact of the transmitted power on data rate:} In Fig.~\ref{CvP}, we evaluated the downlink data rate of MWSP and MWMP scenarios with different transmitted power. We can see that the data rate in the MWSP scenario is higher than that of the fixed single-element antenna baseline, which verifies that relocating the PA close to the user significantly reduces path loss. It can be also observed that the data rate in MWMP scenario is consistently higher than that of the MWSP scenario, indicating that the optimized positions of PAs in MWMP scenario provide higher channel gain. Furthermore, since the PAs deployments in MWMP scenario achieve lower path loss, the data rate in MWMP scenario is higher than that of the fixed antenna scenario. In the MWMP scenario, when the number of PAs is less than 9, the simulated data rate approaches the theoretical maximum, and the achievable rate increases as the number of PAs grows, which indicates that enlarging the array size effectively enhances the channel gain. However, when the number of PAs reaches 9, the simulated data rate becomes lower than that of the case with 7 PAs. This is because increasing the number of PAs leads to larger positioning errors, and thus the relocated PAs deviate from their optimal locations, which reduce the effective channel gain. This finding reveals a trade-off between positioning accuracy and communication performance: Although more PAs theoretically increase data rate, excessive PAs introduces positioning errors and impairs the expected performance.

\begin{figure}[t!]
\centering
\includegraphics[width =3in]{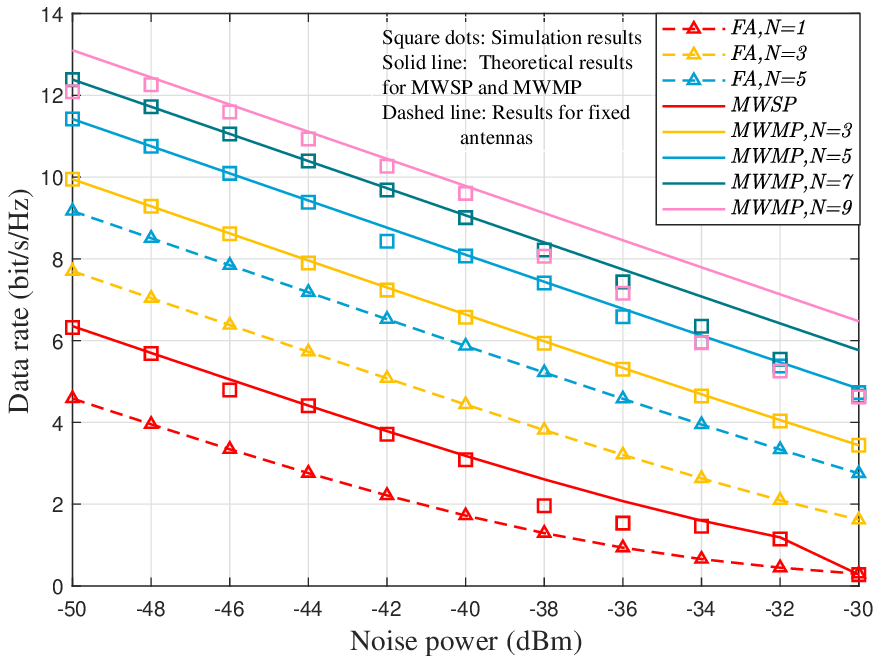}
\caption{Data rate of MWSP and MWMP scenarios versus the noise power, and the transmitted power is set to $P_t$ = 20 dBm.}
\label{CvN}
\end{figure}

\emph{2) Impact of the noise power on data rate:} Fig.~\ref{CvN} illustrates the downlink data rate as the noise increases. Since the transmitted power remains unchanged, increasing the noise power leads to a decrease in SNR, and the data rate decreases as the noise power increases. When the noise power is low, the simulated data rates in both MWSP and MWMP scenarios approach the theoretical rate. The MWMP scenario generally achieves higher data rate than MWSP scenario, and the MWSP scenario achieves a higher data rate than the fixed single-element antenna baseline. The data rates further improve in MWMP scenario as the number of PAs increases, which indicates that the superposition of coherent signals enhances the array gain. However, when the noise power becomes higher, the simulated data rate fails to achieve the theoretical rate when more than 5 PAs are activated, and the data rate decreases as the number of PAs increases. These phenomena indicate that noise affects both positioning accuracy and data rate simultaneously. Moreover, once the positioning accuracy deteriorates, the data rate is further impaired. Therefore, the noise has a serious double-impact on data rate, which confirms our~\textbf{Remark~\ref{remark double impact}}. Owing to positioning errors, the relocated PAs may deviate from the theoretical optimum and fall into suboptimal positions. Nevertheless, the proposed approaches reduce path loss and enhance communication data rate compared with fixed antenna systems.

\section{Conclusion}

In this paper, we started by examining earlier research related to PASS and indoor positioning, and we then proposed the multi-waveguide-based uplink indoor positioning approaches. We considered two possible scenarios, MWSP and MWMP scenarios. For the MWSP scenario, we proposed an RSSI-based ranging approach and developed an LS positioning algorithm. To obtain deeper insights, we performed a detailed error analysis of the MWSP-based LS positioning algorithm. Subsequently, in order to improve the downlink communication performance, the MWMP scenario was further investigated. We derived the closed-form expression of the superposed signal and designed a grid-search positioning algorithm that exploits the signal power. Furthermore, we conducted the error analysis of the proposed positioning algorithm. Then, based on the estimated user position, we relocated the PAs and analyzed the downlink data rate in both MWSP and MWMP scenarios. Future work will investigate the multipath models under more realistic propagation conditions and extend the current positioning framework from 2D to 3D scenarios. In addition, other advanced positioning techniques will be explored to further enhance the system robustness.

\numberwithin{equation}{section}
\section*{Appendix~A: Proof of Lemma~\ref{lemma1:determinant of A and the area}} \label{Appendix:As}
\renewcommand{\theequation}{A.\arabic{equation}}
\setcounter{equation}{0}

We define the vectors ${{\bf{v}}_1} = \left[ {\begin{array}{*{20}{c}}
{\left( {{x_2} - {x_1}} \right)}&{\left( {{y_2} - {y_1}} \right)}&0
\end{array}} \right]$ and ${{\bf{v}}_2} = \left[ {\begin{array}{*{20}{c}}
{\left( {{x_3} - {x_1}} \right)}&{\left( {{y_3} - {y_1}} \right)}&0
\end{array}} \right]$. Then, the cross product of ${{\bf{v}}_1}$ and ${{\bf{v}}_2}$ is given by:
\begin{equation}\label{cross product}
\begin{array}{*{20}{l}}
{{{\bf{v}}_1} \times {{\bf{v}}_2} = \det \left( {\begin{array}{*{20}{c}}
{\bf{i}}&{\bf{j}}&{\bf{k}}\\
{\left( {{x_2} - {x_1}} \right)}&{\left( {{y_2} - {y_1}} \right)}&0\\
{\left( {{x_3} - {x_1}} \right)}&{\left( {{y_3} - {y_1}} \right)}&0
\end{array}} \right)}\\
{ = \left( {\left( {{x_2} - {x_1}} \right)\left( {{y_3} - {y_1}} \right) - \left( {{x_3} - {x_1}} \right)\left( {{y_2} - {y_1}} \right)} \right){\bf{k}}},
\end{array}
\end{equation}
where $\times$ represents the cross product operator. $\bf{i}$, ${\bf{j}}$ and ${\bf{k}}$ denote the basis vectors of the three-dimensional space.

According to the definition of the cross product modulus~\cite{xiandai}, we can obtain:
\begin{equation}\label{cross product modulus}
\begin{aligned}
\left\| {{{\bf{v}}_1} \times {{\bf{v}}_2}} \right\| &= \left| {\left( {{x_2} - {x_1}} \right)\left( {{y_3} - {y_1}} \right) - \left( {{x_3} - {x_1}} \right)\left( {{y_2} - {y_1}} \right)} \right|\\
 &= \left\| {{{\bf{v}}_1}} \right\|\left\| {{{\bf{v}}_2}} \right\|\sin {\theta _v},
\end{aligned}
\end{equation}
where $\sin {\theta _v}$ is the angle between vectors ${{\bf{v}}_1}$ and ${{\bf{v}}_2}$.

Since $\left\| {{{\bf{v}}_1}} \right\|\left\| {{{\bf{v}}_2}} \right\|\sin {\theta _v}$ is twice the area of the triangle formed by PAs, the conclusion in Lemma~\ref{lemma1:determinant of A and the area} can be derived, and the proof is complete.

\numberwithin{equation}{section}
\section*{Appendix~B: Proof of Lemma~\ref{lemma2:cos_i}} \label{Appendix:Bs}
\renewcommand{\theequation}{B.\arabic{equation}}
\setcounter{equation}{0}
We define the direction vector of the waveguide along the x-axis and the signal as ${{\bf{u}}_x}$ and ${{\bf{d}}_x}$, which can be respectively written as:
\begin{equation}\label{waveguide direction vector-x}
{{\bf{u}}_x} = \left[ {\begin{array}{*{20}{c}}
1&0&0
\end{array}} \right],
\end{equation}  
and
\begin{equation}\label{signal direction vector-x}
{{\bf{d}}_x} = \left[ {\begin{array}{*{20}{c}}
{{x_u} - {x_1}}&{{y_u} - {y_1}}&{ - h}
\end{array}} \right].
\end{equation}

Then, the cosine of angle ${\theta_1}$ is derived as follows:
\begin{equation}\label{cosine of the angle-x}
\cos {\theta _1} = \frac{{{{\bf{u}}_x}{{\bf{d}}_x}}}{{\left\| {{{\bf{u}}_x}} \right\|\left\| {{{\bf{d}}_x}} \right\|}} = \frac{{{x_u} - {x_1}}}{{{d_{u1}}}}.
\end{equation}

The analysis of the waveguides along the y-axis and along the diagonal are identical to the waveguide along the x-axis. Similarly, the direction vector of the waveguide along the y-axis and the signal can be respectively given by:
\begin{equation}\label{waveguide direction vector-y}
{{\bf{u}}_y} = \left[ {\begin{array}{*{20}{c}}
0&1&0
\end{array}} \right],
\end{equation}  
and
\begin{equation}\label{signal direction vector-y}
{{\bf{d}}_y} = \left[ {\begin{array}{*{20}{c}}
{{x_u} - {x_2}}&{{y_u} - {y_2}}&{ - h}
\end{array}} \right].
\end{equation}

The cosine of angle between signal and the waveguide along y-axis can be written as:
\begin{equation}\label{cosine of the angle-y}
\begin{aligned}
\cos {\theta _2} = \frac{{{{\bf{u}}_y}{{\bf{d}}_y}}}{{\left\| {{{\bf{u}}_y}} \right\|\left\| {{{\bf{d}}_y}} \right\|}} = \frac{{{y_u} - {y_2}}}{{{d_{u2}}}}.
\end{aligned}
\end{equation}

The direction vector of the waveguide along the diagonal and signal can be separately expressed as:
\begin{equation}\label{waveguide direction vector-m}
{{\bf{u}}_m} = \frac{1}{{\sqrt {D_1^2 + D_2^2} }}\left[ {\begin{array}{*{20}{c}}
{{D_1}}&{{D_2}}&0
\end{array}} \right],
\end{equation}  
and
\begin{equation}\label{signal direction vector-m}
{{\bf{d}}_m} = \left[ {\begin{array}{*{20}{c}}
{{x_u} - {x_3}}&{{y_u} - {y_3}}&{ - h}
\end{array}} \right].
\end{equation}

For the waveguide along the diagonal, the cosine of angle between signal and waveguide can be given by:
\begin{equation}\label{cosine of the angle-m}
\begin{aligned}
\cos {\theta _3} = \frac{{{{\bf{u}}_m}{{\bf{d}}_m}}}{{\left\| {{{\bf{u}}_m}} \right\|\left\| {{{\bf{d}}_m}} \right\|}} = \frac{{{D_1}\left( {{x_u} - {x_3}} \right) + {D_1}\left( {{y_u} - {y_3}} \right)}}{{{d_{u3}}\sqrt {D_1^2 + D_2^2} }}.
\end{aligned}
\end{equation}

Hence, the cosine values of angles can be derived in Lemma~\ref{lemma2:cos_i}, and the proof is complete.

\numberwithin{equation}{section}
\section*{Appendix~C: Proof of Lemma~\ref{lemma3:superimposed received signal closed-form}} \label{Appendix:Cs}
\renewcommand{\theequation}{C.\arabic{equation}}
\setcounter{equation}{0}
We focus on the summation term in \eqref{rewritten superimposed received signal x}. Since ${\psi _i} = \frac{{2\pi d\cos {\theta _i}}}{\lambda }$, the summation term can be written as:
\begin{equation}\label{the summation term}
\begin{aligned}
&\sum\limits_{n =  - \left( {N - 1} \right)/2}^{\left( {N - 1} \right)/2} {\exp \left( {jn{\psi _i}} \right)}  = \exp \left( { - j\frac{{N - 1}}{2}{\psi _i}} \right)\\
 &+ \exp \left( { - j\frac{{N - 3}}{2}{\psi _i}} \right) +  \ldots  + \exp \left( {j\frac{{N - 3}}{2}{\psi _i}} \right)\\
 &+ \exp \left( {j\frac{{N - 1}}{2}{\psi _i}} \right).
\end{aligned}
\end{equation}

By applying the geometric series summation, \eqref{the summation term} can be transformed into:
\begin{equation}\label{the summation term derived into}
\begin{aligned}
&\sum\limits_{n =  - \left( {N - 1} \right)/2}^{\left( {N - 1} \right)/2} {\exp \left( {jn{\psi _i}} \right)}  = \frac{{{e^{ - j\frac{{N - 1}}{2}{\psi _i}}} - {e^{j\frac{{N + 1}}{2}{\psi _i}}}}}{{1 - {e^{j{\psi _i}}}}}\\
 &= \frac{{{e^{j\frac{{{\psi _i}}}{2}}}\left( {{e^{ - j\frac{{N{\psi _i}}}{2}}} - {e^{j\frac{{N{\psi _i}}}{2}}}} \right)}}{{{e^{j\frac{{{\psi _i}}}{2}}}\left( {{e^{ - j\frac{{{\psi _i}}}{2}}} - {e^{j\frac{{{\psi _i}}}{2}}}} \right)}} = \frac{{\sin \left( {\frac{{N{\psi _i}}}{2}} \right)}}{{\sin \left( {\frac{{{\psi _i}}}{2}} \right)}}.
\end{aligned}
\end{equation}

By substituting \eqref{the summation term derived into} into \eqref{rewritten superimposed received signal x}, we can obtain the closed-form expression of the superimposed received signal in \eqref{closed-form expression of the superimposed received signal in lemma2}. The proof is complete.

\bibliographystyle{IEEEtran}
\bibliography{IEEEabrv,Integrated_Positioning_and_Communications_for_PASS}

\end{document}